\documentclass[12pt]{article}
\usepackage{color}
\usepackage[dvips]{graphicx}
\usepackage{dcolumn}
\usepackage{amsmath}
\usepackage{url}

\begin{document}
\Large
\begin{center}
  Diagrammatic Multiplet-Sum Method (MSM) Density-Functional Theory (DFT):
  III. Inclusion of Relaxation and Application to LiH
\end{center}
\normalsize

\vspace{0.5cm}

\noindent
Mark E.\ Casida\\
{\em Laboratoire de Spectrom\'etrie, Interactions et Chimie th\'eorique 
(SITh),
D\'epartement de Chimie Mol\'eculaire (DCM, UMR CNRS/UGA 5250),
Institut de Chimie Mol\'eculaire de Grenoble (ICMG, FR2607), 
Universit\'e Grenoble Alpes (UGA)
301 rue de la Chimie, BP 53, F-38041 Grenoble Cedex 9, FRANCE\\
e-mail: mark.casida@univ-grenoble-alpes.fr} 

\vspace{0.5cm}

\noindent
Abraham Ponra\\
{\em African Institute for Mathematical Sciences (AIMS),
AIMS-Cameroon, P.O.\ Box 608, Limbe, CAMEROON\\
University of Maroua, P.O.\ Box 814, Maroua, CAMEROON\\
e-mail: abraham.ponra@aims-cameroon.org}

\vspace{0.5cm}

\noindent
Gadzikano Munyuki\\
{\em Department of Chemistry and Earth Sciences,
Faculty of Science, University of Zimbabwe,
Harare, ZIMBABWE\\
e-mail: gadzie@gmail.com}

\vspace{0.5cm}

\noindent
Bharathi Natarajan\\
{\em Independent Researcher, Coimbatore,
Tamil Nadu, INDIA\\
e-mail: phyever@gmail.com}

\vspace{0.5cm}



\begin{center}
{\bf Abstract}
\end{center}

Ideal density-functional approximations (DFAs) should account for dynamic, 
static, and nondynamic correlation. While common DFAs struggle with the 
latter two, the Ziegler-Rauk-Baerends-Daul multiplet sum method (MSM) 
provides a pragmatic way to include static correlation. In this article, 
we use diagrammatic MSM density-functional theory (diag MSM DFT) using 
the two-orbital 
two-electron model (TOTEM) to extend MSM DFT to include nondynamic 
correlation without relying on symmetry arguments. Building on previous 
formulations 
[A.\ Ponra, C.\ Bakasa, A.J.\ Etindele, and M.E.\ Casida, 
{\em J.\ Chem.\ Phys.}\ {\bf 159}, 244306 (2023); 
M.E.\ Casida, A.\ Ponra, C.\ Bakasa, and A.J.\ Etindele,
{\em J.\ Chem.\ Phys.}\ {\bf 162}, 144317 (2025)]
that lacked relaxation effects, this article incorporates relaxation via 
nonorthogonal configuration interaction (NOCI). We demonstrate that this 
modified diag MSM DFT produces an accurate ground-state potential energy 
curve (PEC) for lithium hydride (LiH), even at the 
ionic-to-open-shell-singlet avoided crossing characterized by 
significant charge transfer. This encouraging result suggests that the
model can be extended to (at least) other singly and multiply-bonded diatomic 
molecules, while providing insight into a novel way to include strong
correlation in DFT.

\section{Introduction}
\label{sec:intro}

Modern density-functional theory (DFT) is an exact formalism defined
upon the two Hohenberg-Kohn theorems \cite{HK64} and the Kohn-Sham (KS)
reformulation \cite{KS65}.  The exact exchange-correlation (xc)
functional $E_{xc}[\rho]$ is the target for designing density-functional
approximations (DFAs). A similar statement may be made about meta generalized 
gradient approximations or hybrid functionals which, strictly speaking,
fall into the category of generalized KS theory \cite{SGV+96}.  Both in 
the original and in the generalized theories, the functional $E_{xc}$
is a complicated beast, containing information about the difference between
the kinetic energy of the real interacting system of electrons and
the fictitious KS system of noninteracting electrons, exchange, and
correlation.  Electron correlation may be classified in a somewhat fuzzy, 
albeit useful manner, as dynamic, static, and nondynamic \cite{BS94}.
Dynamic correlation applies when a single determinant (SDET)
is a good first
approximation to the interacting wave function.  Static (degenerate) 
correlation arises due to degeneracies among  SDET states 
which normally implies the presence of symmetry.  Nondynamic (quasidegenerate) 
correlation arises due to quasidegenerate states and is often
associated with bond making and bond breaking.  Evidently, {\em getting
nondynamic correlation right is very important for studying 
chemical reactions}.  Both static and nondynamic correlation constitute 
examples of strong correlation.  These are known
to show up as a particle number derivative discontinuity in 
$E_{xc}$ \cite{PPLB82,SS83}
and, in the exact xc potential,
$v_{xc}[\rho](1) = \delta E_{xc}[\rho]/\delta \rho(1)$ 
as peculiar plateaus around atoms and peaks between atoms when bonds 
dissociate \cite{P85,GB96,TMM09}.  Such features are generally absent in
commonly used DFAs and will be very difficult to include when designing
better DFAs.  However commonly used DFAs are known to do a good job at
describing dynamic correlation, but not strong correlation.  Problems
show up in the form of symmetry breaking which leads to triplet instabilities
when calculating excitation energies using response theory (see, e.g.,
Ref.~\cite{CGG+00})
and in the lowest unoccupied molecular orbital (LUMO or L)
falling lower in energy than the highest occupied molecular orbital 
(HOMO or H) energy (see, e.g., Ref.~\cite{TTR+08}).
Our idea is to extend further some classic physical approaches in DFT to 
include strong correlation beyond what these methods have been able to do
so far.  Our strategy is to learn from simple molecules and then to gradually
extend our ideas to more and more complicated cases.  Our hope is to keep the 
evolving method as simple as possible for as long as we can during the 
development process, while simultaneously attaining quantitative results.

It seems appropriate in a volume honoring Axel Becke, to mention that Axel
once said that, ``The focus of ground-state DFA developers has shifted to 
[what he hoped] is the last frontier, `strong' correlation''
\cite{B2014}.  Axel advocated approaching strong correlation in a very 
different way than we are proposing here.  Axel's preference was to try to
find ways to include strong correlation through increasingly accurate 
xc holes.  Our approach is philosophically closer to when
Axel writes in the same article that, ``The ultimate conclusion of 
combining DFT and [wave function theory] is what Bartlett calls \cite{B2010} 
`{\em ab initio} DFT,' a mathematical and theoretical framework that exploits
the best of both worlds and seeks a systematic route to improving both.''
This systematic route can be very complicated and, at least for now, we are 
trying to keep our approach as simple as possible, even if there are
definite philosophical overlaps with Bartlett's approach.  In fact, a major
point for us is to build practical experience with combining DFT
and wave function theory ideas, first for simple systems, and then for 
increasingly complicated systems.

In the late 1970s, Ziegler, Rauk, and Baerends
introduced the multiplet sum method (MSM) to add static correlation
into DFT in an approximate way on the basis of a zero-order guess for the 
wave function \cite{ZRB77}, and this method continues to be widely used
in DFT for estimating the energy of the lowest open-shell singlet state by
using spin symmetry.  Spatial symmetry may also be used \cite{D94,PEMC21}.
It is important to understand that MSM-DFT is {\em not} a route towards
better DFAs, but rather is a way to profit from the strengths of existant
DFAs while still including strong correlation.  The strength of MSM-DFT is
in its simplicity and ease of use, but it does not describe nondynamic
correlation.  Diagrammatic (diag) MSM-DFT was introduced recently as a way to make
educated guesses of DFT matrix elements in a small configuration 
interaction (CI) problem (Article {\bf I} \cite{PBEC23} and 
Article {\bf II} \cite{CPBE25}).  
It is important to understand that here, as in most CI calculations, the
emphasis is on the quality of the energy of the lowest energy state with 
the introduction of ``excited states'' as a way to improve the lowest
energy state.  Nevertheless excited- and ground-states interact via
avoided crossings and conical intersections, so we also find it interesting
to examine the behavior of the excited-state solutions from our 
diag MSM DFT calculations.  So far, emphasis has been on the
two-level two-electron model (TOTEM). 
Article {\bf II} showed promising results for calculating off-diagonal
coupling elements key for describing the 
[Li$^+$  H:$^-$]/[Li$\uparrow$ H$\downarrow$ $\leftrightarrow$
Li$\downarrow$ H$\uparrow$] avoided crossing but also made it clear that
the method is not yet quantitative.  One problem with the diag
MSM DFT description of the TOTEM is that it needs to include orbital
relaxation effects in order to keep the CI expansion as small as possible.
In the present article, we gradually add this missing relaxation through
the use of configuration state functions (CSFs) built from
different relaxed molecular orbitals (MOs).  Many variants
on our basic model have been tested but only the most useful is presented
as well as some of the intermediate results that have helped us to understand
how the different variants behave.

This article is organized as follows: The next section (Sec.~\ref{sec:exact})
presents our choice of comparison data and discusses the principle features
of the avoided crossings seen in the PECs of LiH.  The chemical physics
would seem to be adequately understood that we should be able to build a
simple hybrid DFT/wave function theory (WFT) to be able to 
describe the lowest energy singlet PEC (i.e., for the $X\,^1\Sigma$ state).
Section~\ref{sec:theory} provides a brief review of 
diag MSM MDFT in order to keep the present article relatively
self contained.  The new theory tested in the present article is the
nonorthogonal configuration interaction (NOCI) method using relaxed orbitals
presented in Sec.~\ref{sec:theory1}.
Results are given in Sec.~\ref{sec:results} and the concluding discussion
is given in Sec.~\ref{sec:conclude}.  Additional information about
author contributions, computational details, and a discussion of the origin
of a key approximation made in the present work may be
found in the supplementary information (SI).

%
%

\section{Comparison Data}
\label{sec:exact}

Comparisons will be made against the same high-quality LiH PECs \cite{GL06}
as in 
Articles {\bf I} and {\bf II} which we will label as EXACT.
``Exact'' is, of course, a relative term, but these EXACT curves are
essentially exact with respect to our model calculations and so provide
an excellent comparison.

\begin{figure} 
\begin{center}
\includegraphics[width=0.60\textwidth]{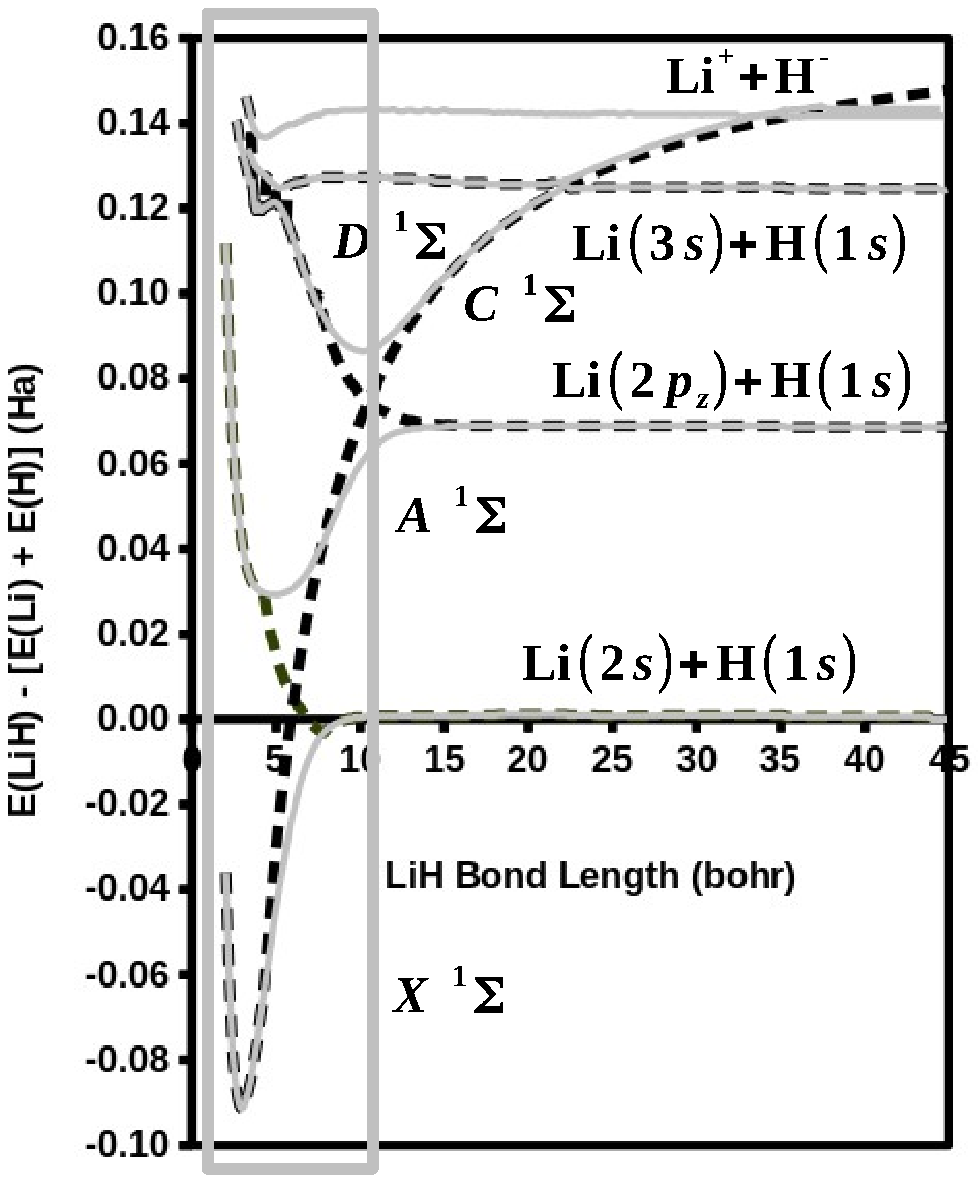} 
\end{center}
\caption{
EXACT singlet PECs for LiH.  The grey adiabatic curves were digitized 
from Fig.~1 of Ref.~\cite{GL06} using {\sc WebPlotDigitizer} 
\cite{WebPlotDigitizer}. The adiabatic curve labels $X\,^1\Sigma$, 
$A\,^1\Sigma$, $C\,^1\Sigma$, and $D\,^1\Sigma$ are traditional.
The dashed black curves are rough diabatic curves more or less constructed
by hand.  A notable exception is the Li$^+$ + H$^-$ diabatic curve which
has the form $E-1/R$ where $R$ is the LiH bond distance and $E$ = 0.17 Ha
and which provides a very nice and rigorous diabatic curve over a large
range of $R$.  The value $E$ is close to the expected value, which is 
the energy needed to transfer an electron Li + H $\rightarrow$ 
Li$^-$ + H$^+$ at infinite $R$.  Specifically, $E$ = IP(Li)-EA(H) = 0.20 Ha 
using the known ionization potential (IP) of the lithium atom and the 
known electron affinity (EA) of the hydrogen atom.
\label{fig:diabats}
}
\end{figure}
Lithium hydride is analogous to sodium hydride (Na$^+$)(H$^-$), a well-known
reagent in organic chemistry.  It is ionically bonded at its equilibrium
geometry but dissociates to the neutral atoms in the gas phase because of
a series of avoided crossings.  These avoided crossings are nicely shown
in {\bf Fig.~\ref{fig:diabats}}.  The Li$^+$ + H$^-$ diabatic curve
crosses the curves that correspond to the Heitler-London valence-bond 
(VB) description of covalent bonding 
[Li-H] = 
[Li$\uparrow$ H$\downarrow$ $\leftrightarrow$ Li$\downarrow$ H$\uparrow$]
where H dissociates to its ground state but Li is in an excited state.
[We will often indicate Lewis dot structures (LDS) by
square brackets when additional clarity seems desirable.]

\begin{figure}  
\begin{center}
\includegraphics[width=0.85\textwidth]{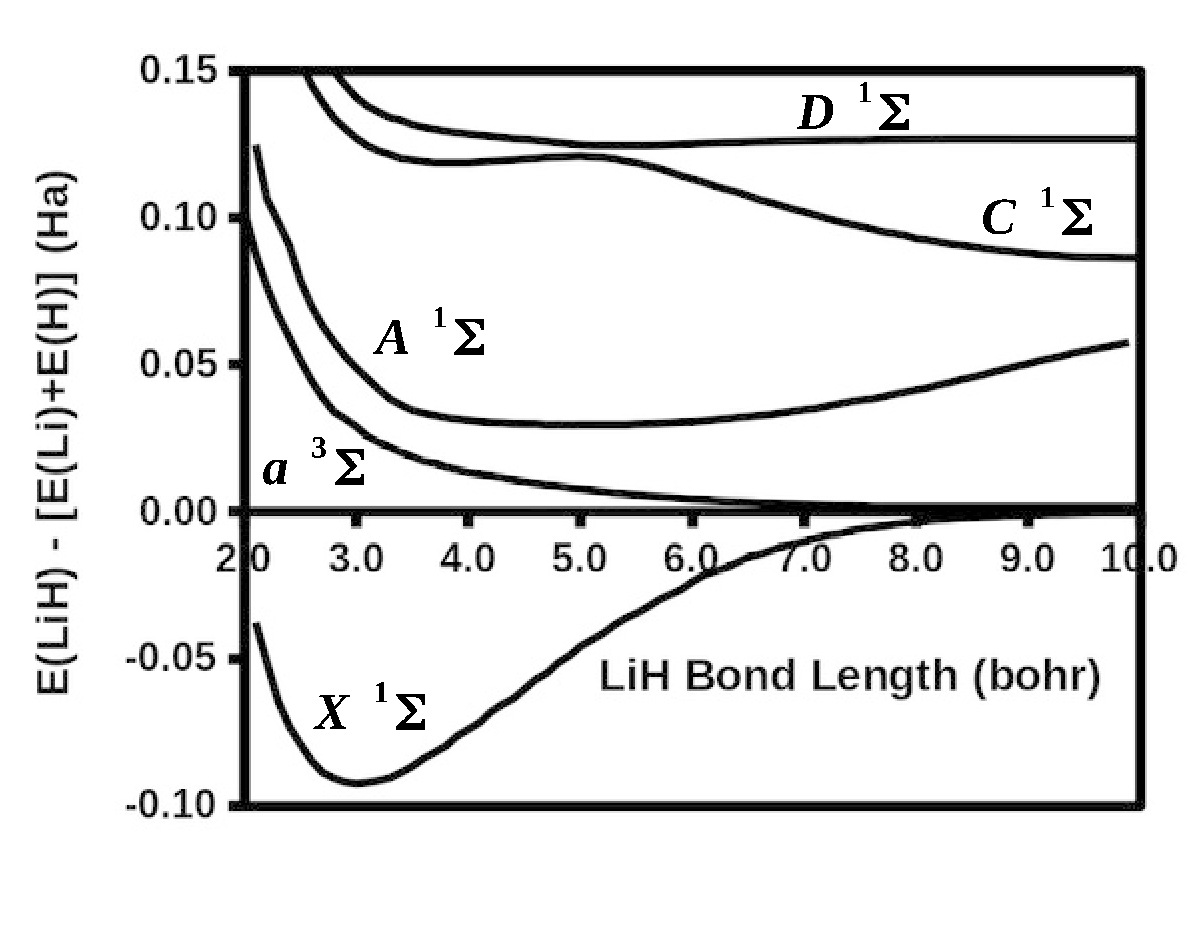} 
\end{center}
\caption{
EXACT PECs for LiH from Ref.~\cite{GL06}, including the lowest triplet
PEC.  (Small wiggles in the curves are from the digitization process using 
{\sc WebPlotDigitizer} \cite{WebPlotDigitizer}.)
Note that the important $A \,^1\Sigma$ state was inadvertently omitted from
the corresponding figures in Articles {\bf I} and {\bf II}.
\label{fig:EXACT}
}
\end{figure}
The present article focuses on the region 2.1 bohr $< R <$ 10.0 bohr shown
in {\bf Fig.~\ref{fig:EXACT}} and, in particular,  on the [Li$^+$ + 
H$^-$]/[Li($2s$) + H($1s$)] ($X\,^1\Sigma$/$A\,^1\Sigma$) avoided crossing.
However it is clear that there is also a contribution from the [Li$^+$ + 
H$^-$]/[Li($2p$) + H($1s$)] ($A\,^1\Sigma$/$C\,^1\Sigma$) near 10 bohr.
As seen in Fig.~3 of Article {\bf II}, the Coulson-Fischer point where
the different-orbitals-for-different-spins (DODS) 
calculation falls below the same-orbitals-for-different-spins 
(SODS) calculation is at about 6.1 bohr
which is (not unexpectedly) also very close to where the 
[Li$^+$ + H$^-$]/[Li($2s$) + H($1s$)] diabatic curves cross.

\begin{figure} 
\begin{center}
\includegraphics[width=1.0\textwidth]{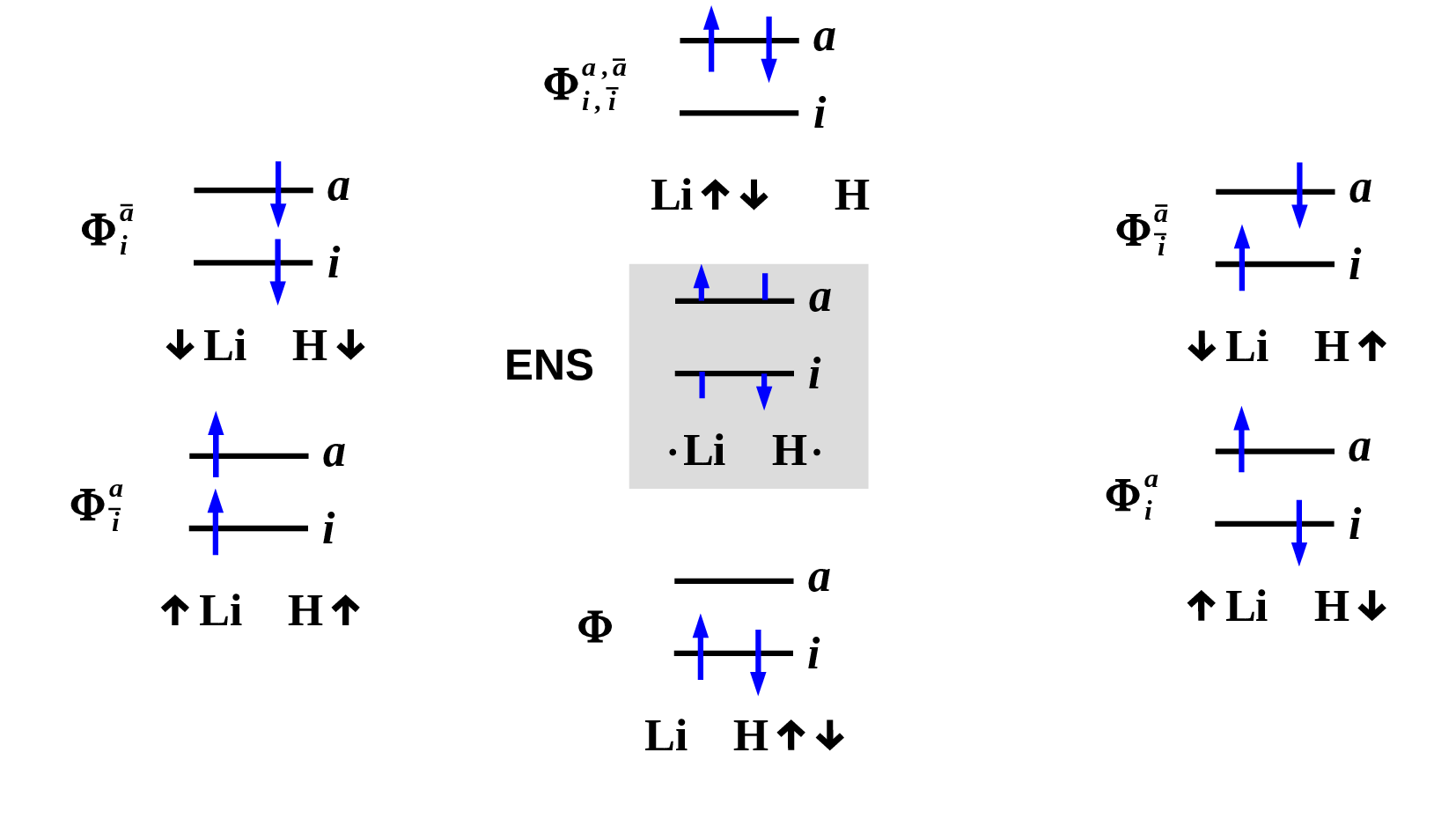}
\end{center}
\caption{Two-level two-electron model (TOTEM) of LiH.  Here
$\Phi = \vert i , {\bar i} \vert$ is a single determinant and
$\Phi_{i,j,...}^{a,b,...} = \cdots b^\dagger j a^\dagger i \Phi$ is
an excited determinant.  The Lewis dot structures (LDS) are only
intended to be suggestive as the spins can delocalize onto both atoms.
\label{fig:TOTEM}
}
\end{figure}
Although not exact, this is an appropriate place to introduce the TOTEM 
that we have used in Articles {\bf I}
and {\bf II} and which is used in the present article.  The most important
and difficult aspect of any multideterminant (MDET) 
approach is the choice of an orbital reference space.  
This is where ``chemical intuition'' (i.e., historical and personal practical
experience) intervenes.  The usual textbook molecular orbital (MO) diagram for
LiH involves a HOMO  bonding combination
that is mostly located on the H but has some component on Li and a 
LUMO  antibonding combination that is 
mostly located on the Li but has some component on H (see the SI of 
Article {\bf I}).  This is confirmed by {\sc Molden} \cite{molden}
images of these orbitals shown in Article {\bf II} and in the corresponding
SI.  This gives us the TOTEM model illustrated in {\bf Fig.~\ref{fig:TOTEM}}.
{\em A priori} this allows us to describe the [Li($2s$) + H($1s$)]
and [Li$^+$ + H:$^-$] diabatic curves, which should be enough to describe
the avoided crossing of principal interest (but not the 
[Li$^+$ + H$^-$]/[Li($2p_z$)+H($1s$)] avoided crossing!)  The question
also arises in this particular TOTEM as to the energy of the doubly excited
[Li:$^-$ + H$^+$] state.  In the asymptotic (large $R$) limit, we expect
an energy equal to IP(H)-EA(Li) = 0.48 Ha on the basis of experimental values.
For comparison, the relaxed energy at $R$ = 10.0 bohr of the doubly excited
SDET is 0.35 Ha.  This is an important
number to keep in mind when examining the results of our calculations.

%
%

\section{Diagrammatic MSM-DFT}
\label{sec:theory}

This section primarily contains provides a brief review in order to keep the
present work as self-contained as possible.
The basic idea for diagrammatic MSM-DFT has already been presented
in Article {\bf I} and was completed---except for relaxation---in 
Article {\bf II} for the TOTEM.  Here we give a minimum of review
and describe how relaxation is to be included.  Since our ultimate
objective (which is beyond the scope of this article) is to get 
excited states by doing response theory on the ground state 
(GS), and because symmetry breaking is known to lead to
triplet instabilities in response theory, we insist upon using a
SODS theory.


\begin{table}
\begin{center}
\begin{tabular}{cc}
\hline \hline
Number & Function \\
\hline
\multicolumn{2}{c}{Single Determinantal}\\
1 & $\Phi_0 = \Phi = \Phi_{\mbox{\tiny GS}}$ \\
2 & $\Phi_M = \Phi_i^a = \Phi_{\mbox{\tiny MIX}}$ \\
3 & $\Phi_{\bar M} = \Phi_{\bar i}^{\bar a}$ \\
4 & $\Phi_D = \Phi_{i,{\bar i}}^{a,{\bar a}} = \Phi_{\mbox{\tiny DEX}}$ \\
\multicolumn{2}{c}{Triplet CSF}\\
  & $\Phi_T = (1/\sqrt{2})(\Phi_{\bar i}^{\bar a}-\Phi_i^a)$ \\
\multicolumn{2}{c}{Singlet CSFs}\\
1 & $\Phi_0 = \Phi = \Phi_{\mbox{\tiny GS}}$ \\
2 & $\Phi_S = (1/\sqrt{2})(\Phi_{\bar i}^{\bar a}+\Phi_i^a) 
= \Phi_{\mbox{\tiny OSS}}$ \\
3 & $\Phi_D = \Phi_{i,{\bar i}}^{a,{\bar a}} = \Phi_{\mbox{\tiny DEX}}$ \\
\hline \hline
\end{tabular}
\end{center}
\caption{\label{tab:CSF} Single-determinantal and spin-adapted 
2-electron configuration state functions (CSFs) for $M_S=0$.
See also Fig.~\ref{fig:TOTEM}.  GS stands for the ``ground-state'' SDET;  
T is the triplet CSF; MIX is a mixed-symmetry SDET;  OSS is the 
``open-shell singlet'' CSF; and DEX is the ``doubly-excited'' SDET.
}
\end{table}

All of the work up to this point has assumed a single set of MOs
obtained from some single reference (REF)
calculation. We will return to this choice of REF later.
For now, let us assume that this has already been done.  
We first consider all the determinants with $M_S=0$
and then make the spin-symmetry adapted CSFs
whose labelling is given in {\bf Table~\ref{tab:CSF}}.  
The triplet state energy is,
\begin{equation}
  E^{\mbox{\tiny REF}}_T = E_{KS}[\Phi_{\bar i}^a] \, ,
  \label{eq:theory.1}
\end{equation}
where we have expressed the KS energy as a functional of the 
KS SDET wave function, in this case $\Phi_{\bar i}^a$
constructed from the REF MOs.  
The MSM singlet wave function has the form,
\begin{equation}
  \Psi = C_{\mbox{\tiny GS}} \Phi^{\mbox{\tiny REF}}_{\mbox{\tiny GS}} 
   + C_{\mbox{\tiny OSS}} \Phi^{\mbox{\tiny REF}}_{\mbox{\tiny OSS}} 
   + C_{\mbox{\tiny DEX}} \Phi^{\mbox{\tiny REF}}_{\mbox{\tiny DEX}} \, ,
  \label{eq:theory.2}
\end{equation}
which leads to the small CI equation,
\begin{equation}
  \left[ \begin{array}{ccc} E^{\mbox{\tiny REF}}_0 & \sqrt{2}D & B \\
                            \sqrt{2}D & E_M^{\mbox{\tiny REF}} + A & \sqrt{2} C \\
                            B & \sqrt{2} C & E^{\mbox{\tiny REF}}_D \end{array} \right]
  \left( \begin{array}{c} C_{\mbox{\tiny GS}} \\ C_{\mbox{\tiny OSS}} \\ C_{\mbox{\tiny DEX}} \end{array} \right)
  = E \left( \begin{array}{c} C_{\mbox{\tiny GS}} \\ C_{\mbox{\tiny OSS}} \\ C_{\mbox{\tiny DEX}} \end{array} \right)
  \, . 
  \label{eq:theory.3}
\end{equation}
The matrix elements $A$, $B$, $C$, and $D$ were derived in previous articles
({\bf Table~\ref{tab:articles}}) but are reviewed here for the sake of 
completeness.
In Eq.~(\ref{eq:theory.3}), $\Phi_{\mbox{\tiny GS}}=\Phi_0$ designates the ground-state (GS) SDET,
$\Phi_{\mbox{\tiny OSS}}=\Phi_S$ designates the open-shell singlet (OSS)
CSF, and $\Phi_{\mbox{\tiny DEX}}=\Phi_D$ designates the doubly-excited 
(DEX) SDET.
The SDET energies are calculated as,
\begin{equation}
  E^{\mbox{\tiny REF}}_0 = E[\Phi_0] \,\,\,\, , \,\,\,\, 
  E^{\mbox{\tiny REF}}_M = E[\Phi_M] = E[\Phi_{\bar M}] \,\,\,\, , \,\,\,\, 
  E^{\mbox{\tiny REF}}_D = E[\Phi_D] \, ,
  \label{eq:theory.4}
\end{equation}
using the REF MOs.  Following Ref.~\cite{ZRB77}, 
\begin{equation}
  A = E^{\mbox{\tiny REF}}_M - E^{\mbox{\tiny REF}}_T \, .
  \label{eq:theory.5}
\end{equation}
From Article {\bf I},
\begin{equation}
  B = A \, .
  \label{eq:theory.6}
\end{equation}
How $C$ and $D$ are calculated depends upon the choice of REF.


\begin{table}
\begin{center}
\begin{tabular}{ll}
\hline \hline
Article &               \\
\hline
{\bf I} \cite{PBEC23} & Determination of A via MSM-DFT and of B via diag MSM-DFT \\
{\bf II} \cite{CPBE25} & Determination of C and D via diag MSM-DFT \\
{\bf III} & Present work \\
\hline \hline
\end{tabular}
\end{center}
\caption{\label{tab:articles} 
Summary of key articles in diag MSM-DFT.
}
\end{table}


Both Articles {\bf I} and {\bf II} used a REF calculation with half an 
electron of each
spin in orbitals $i$ and $a$.  This choice was made, in part, to minimize 
symmetry breaking so that we could work with SODS orbitals. We will refer 
to this as the ensemble (ENS) reference as the density matrix may be 
obtained by constructing a two-electron ENS in various ways and then forming 
the corresponding one-electron reduced density matrix,
\begin{eqnarray}
  \hat{\gamma}_{\mbox{\tiny ENS}} 
  & = & \frac{1}{2} \left( \vert i \rangle \langle i \vert + 
        \vert {\bar i} \rangle \langle {\bar i} \vert +
        \vert a \rangle \langle a \vert + \langle {\bar a} \rangle 
        \langle {\bar a} \vert \right) \nonumber \\
  & = & \frac{1}{4} \left( \hat{\gamma}_0 + \hat{\gamma}_S + \hat{\gamma}_T
    + \hat{\gamma}_D \right) \nonumber \\
  & = & \frac{1}{2} \left( \hat{\gamma}_0 + \hat{\gamma}_D \right) \nonumber \\
  & = & \frac{1}{2} \left( \hat{\gamma}_M + \hat{\gamma}_{\bar M} \right)
    =  \frac{1}{2} \left( \hat{\gamma}_S + \hat{\gamma}_T \right)
        \nonumber \\
  & = & \hat{\gamma}_S = \hat{\gamma}_T \, .
  \label{eq:theory.7}
\end{eqnarray}
We may now specify more clearly that we will always be taking,
\begin{eqnarray}
  B = A = E^{\mbox{\tiny ENS}}_M - E^{\mbox{\tiny ENS}}_T \, .
  \label{eq:theory.7.5}
\end{eqnarray}
Each line provides a different justification for choosing REF = ENS. 
For example, line 2 of Eq.~(\ref{eq:theory.7}) tries to
be unbiased.  Line 3 of Eq.~(\ref{eq:theory.7}) suggests that a compromise
is being made between the ground and doubly-excited CSF.  Line 4 of 
Eq.~(\ref{eq:theory.7}) indicates that this is an especially good choice
for describing the mixed-symmetry or triplet-and-open-shell space.
Line 5 of Eq.~(\ref{eq:theory.7}) further emphasizes that these orbitals
are a good choice for either the triplet or the open-shell energy calculation.
This was indeed found for the triplet state in Articles {\bf I} and {\bf II}.

\begin{figure}  
\begin{center}
\includegraphics[width=0.70\textwidth]{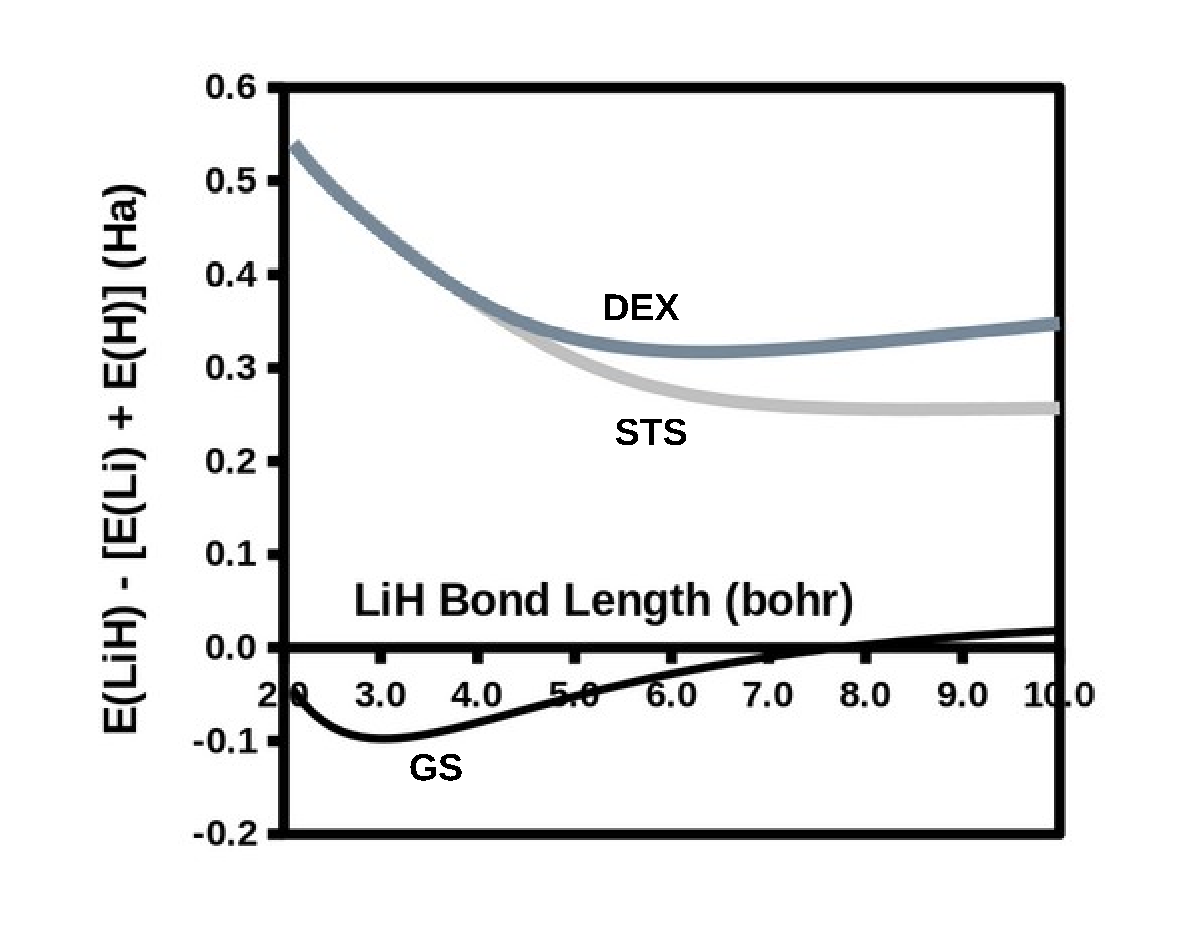} 
\end{center}
\caption{
Relaxed GS SDET PEC, relaxed DEX SDET PEC, and a PEC obtained by
adding Slater's transition state (STS) excitation energy 
to the relaxed GS SDET PEC.
\label{fig:LiH_STS_v3}
}
\end{figure}
Perhaps just as importantly is the fact that occupying the LUMO gives it
some {\em physical meaning} that might otherwise be absent.  
Axel Becke called this the OOO principle, for ``occupied orbitals only'' \cite{B2014}.
Such physical
meaning is important when trying to construct compact wave function
representations.  In particular,
exciting half an electron of each spin from the HOMO to the LUMO is 
exactly Slater's transition
state (STS) for the HOMO$^2$ $\rightarrow$ LUMO$^2$ double 
excitation. The STS for making a double excitation is,
\begin{eqnarray}
  \Delta E_D & = & E_D - E_0 \nonumber \\
   & = & E(n_H^\alpha=1, n_H^\beta=1, n_L^\alpha=0, n_L^\beta=0) \nonumber \\
   & = & \int_0^1 \frac{E(n_H^\alpha=\lambda, n_H^\beta=\lambda, 
         n_L^\alpha=1-\lambda, n_L^\beta=1-\lambda)}{d\lambda} \, d\lambda \nonumber \\
   & = & \int_0^1 \left( \epsilon_L^\alpha(n_H^\alpha=\lambda, n_H^\beta=\lambda, 
         n_L^\alpha=1-\lambda, n_L^\beta=1-\lambda)  \right. \nonumber \\
         & + & \epsilon_L^\beta(n_H^\alpha=\lambda, n_H^\beta=\lambda, 
         n_L^\alpha=1-\lambda, n_L^\beta=1-\lambda) \nonumber \\ 
         & - & \epsilon_H^\alpha(n_H^\alpha=\lambda, n_H^\beta=\lambda, 
         n_L^\alpha=1-\lambda, n_L^\beta=1-\lambda) \nonumber \\
         & - & \left. \epsilon_H^\beta(n_H^\alpha=\lambda, n_H^\beta=\lambda, 
         n_L^\alpha=1-\lambda, n_L^\beta=1-\lambda)
         \right) \, d\lambda \nonumber \\
   & \approx & 2\left( \epsilon_L(ENS) - \epsilon_H(ENS) \right) \, .
  \label{eq:3.1}
\end{eqnarray}
This is the STS 
line in {\bf Fig.~\ref{fig:LiH_STS_v3}}.
Also shown is the DEX PEC calculated with the variationally optimized
DEX orbitals. 
We can calculate this because {\sc deMon2k}
has a maximum-overlap orbital-following algorithm that makes sure that
the correct orbitals are occupied during the iterations in order to
avoid variational collapse.  Nevertheless, two different SCF solutions
were found and only the lowest energy one was used.  This DEX PEC 
solution is the closest thing that we have to the exact DEX PEC.
As might be expected based upon the variational principle,
the DEX PEC calculated with the ENS MOs (not shown) 
is higher in energy but parallel to the DEX 
curve.  Theory suggests that the STS 
line should be an approximation to the exact 
DEX PEC.  As we can see, this STS approximation works well for $R < 4.0$ bohr,
but thereafter is only a very rough approximation.

What about the $C$ and $D$ matrix elements?  Article {\bf II} postulated that 
\begin{eqnarray}
  C = f^{\mbox{\tiny ENS}}_{i,a}[\gamma_{\mbox{\tiny DEX}}^{\mbox{\tiny ENS}}] \nonumber \\
  D = f^{\mbox{\tiny ENS}}_{i,a}[\gamma_{\mbox{\tiny GS}}^{\mbox{\tiny ENS}}] \, ,
  \label{eq:theory.8}
\end{eqnarray}
on the basis of similarities with wave function theory.  Here
$f^{\mbox{\tiny ENS}}_{i,a}$ is the off-diagonal element of the KS orbital
hamiltonian matrix calculated in the ENS MO basis set.
However this choice is not unique because another seemingly logical choice 
for the reference is the GS.  For the $C$ and $D$ matrix elements 
(temporarily denoted as $\tilde C$ and $\tilde D$ for consistency with the
SI), 
\begin{eqnarray}
  \tilde C & = &  f^{\mbox{\tiny GS}}_{i,a}[\gamma_{\mbox{\tiny DEX}}^{\mbox{\tiny GS}}]
     \approx 2 f^{\mbox{\tiny ENS}}_{i,a}[\gamma_{\mbox{\tiny DEX}}^{\mbox{\tiny ENS}}] = 2C \nonumber \\
   \tilde D & = & f^{\mbox{\tiny GS}}_{i,a} 
          [\gamma_{\mbox{\tiny GS}}^{\mbox{\tiny GS}}]  = 0 \, .
  \label{eq:theory.9}
\end{eqnarray}
where 
$f^{\mbox{\tiny GS}}_{i,a}[\gamma_{\mbox{\tiny DEX}}^{\mbox{\tiny GS}}]$ is the off-diagonal element of the KS orbital hamiltonian matix in 
the GS MO basis set calculated with the DEX matrix element constructed with 
DEX occupation numbers but GS MOs,
$f^{\mbox{\tiny ENS}}_{i,a}[\gamma_{\mbox{\tiny DEX}}^{\mbox{\tiny ENS}}]$
is the off-diagonal element of the KS orbital hamiltonian matrix 
in the ENS MO basis set calculated with the approximate DEX density matrix 
constructed with DEX occupation numbers but ENS MOs, and 
$f^{\mbox{\tiny GS}}_{i,a}$ is the off-diagonal element of the KS orbital
hamiltonian matrix in the GS MO basis set calculated with the GS density
matrix constructed with GS occupation numbers.  The second line of 
Eq.~(\ref{eq:theory.9}) is the Kohn-Sham equivalent of Brillouin's 
theorem (i.e., the Kohn-Sham orbital matrix is diagonal in the canonical
Kohn-Sham orbitals).  However the approximation in the first line of 
Eq.~(\ref{eq:theory.9}), which involves a change of reference orbitals,
is a little involved to derived.  It is presented diagrammatically
in Article {\bf II} [Compare Fig.~5 and
Eq.~(23) of Article {\bf II}] and an explicit argument
is given in the SI for the present work.

\section{Diagrammatic MSM-DFT with Relaxation}
\label{sec:theory1}



This section contains the new theory studied in this article.  
Specifically, it was recognized in Article {\bf II}, that the 
diagrammatic MSM-DFT method presented thus far is missing important 
relaxation effects.  In principle, these can be included by using
nonorthogonal configuration interaction (NOCI) where
the different CSFs in the CI expansion are obtained from a different
reference.  {\em A priori}, our ideal NOCI wave function should take the form,
\begin{equation}
  \Psi = C_{\mbox{\tiny GS}} \Phi^{\mbox{\tiny GS}}_{\mbox{\tiny GS}} 
   + C_{\mbox{\tiny OSS}} \Phi^{\mbox{\tiny ENS}}_{\mbox{\tiny OSS}} 
   + C_{\mbox{\tiny DEX}} \Phi^{\mbox{\tiny DEX}}_{\mbox{\tiny DEX}} \, ,
  \label{eq:theory.10}
\end{equation}
where the superscripts emphasize that 
$\Phi^{\mbox{\tiny GS}}_{\mbox{\tiny GS}}$ 
is variationally optimal for the ground SDET state and 
$\Phi^{\mbox{\tiny DEX}}_{\mbox{\tiny DEX}}$ is variationally optimal
for the doubly excited SDET state.  We will keep the ENS orbitals for
constructing the open-shell singlet CSF.  It is actually interesting
to not go immediately to three different reference states but to also
look at some intermediate combinations where, say, 
$\Phi^{\mbox{\tiny ENS}}_{\mbox{\tiny OSS}}$ is replaced with 
$\Phi^{\mbox{\tiny GS}}_{\mbox{\tiny OSS}}$ as will be shown in
Sec.~\ref{sec:results}.

Strictly speaking, LiH is a four-electron problem, not a two-electron problem.
However there is ample experience reported in the literature that relaxation
of the the doubly occupied core $1s$ electrons on Li may be neglected to a 
first approximation.  This is clearly an approximation where some care is 
in order.  In particular, care should be taken when applying the TOTEM to 
larger molecules with more complicated electronic configurations, such as 
NaF, where core relaxation may still be a reasonable approximation but 
where valence orbitals outside the TOTEM space are expected to show 
significant relaxation effects which will have to be included in our model.  
However, this is beyond the scope of the present work.

NOCI involves new challenges which we partially address and partially
avoid in our simplified treatment.  Ordinary single reference CI 
involves solving Eq.~(\ref{eq:theory.3}),
\begin{equation}
  {\bf H} \vec{C}_I = E_I \vec{C}_I \, .
  \label{eq:theory.11}
\end{equation}
A multireference relaxed treatment means that there is a CSF overlap
matrix ${\bf S}$ which differs from unity.  Its calculation is tractable
because of the theorem that the overlap matrix of determinants is the 
determinant of the overlap matrices,
\begin{equation}
  \langle i_1, i_2, \cdots, i_N \vert j_1, j_2, \cdots, j_N \rangle
  = \det \left|
  \begin{array}{cccc}
  \langle i_1 \vert j_1 \rangle & \langle i_1 \vert j_2 \rangle &
  \cdots & \langle i_1 \vert i_N \rangle \\
  \langle i_2 \vert j_1 \rangle & \langle i_2 \vert j_2 \rangle &
  \cdots & \langle i_2 \vert i_N \rangle \\
  \vdots & \vdots & \ddots & \vdots \\
  \langle i_N \vert j_1 \rangle & \langle i_N \vert j_2 \rangle &
  \cdots & \langle i_N \vert i_N \rangle 
  \end{array}
  \right| \, .
  \label{eq:theory.19}
\end{equation}
The necessary overlap matrix may then be calculated as,
\begin{eqnarray}
  {\bf S} & = &  \left[
  \begin{array}{ccc}
  \langle \Psi^{\mbox{\tiny GS}}_{\mbox{\tiny GS}} \vert \Psi^{\mbox{\tiny GS}}_{\mbox{\tiny GS}} \rangle &
  \langle \Psi^{\mbox{\tiny GS}}_{\mbox{\tiny GS}} \vert \Psi^{\mbox{\tiny ENS}}_{\mbox{\tiny OSS}} \rangle &
  \langle \Psi^{\mbox{\tiny GS}}_{\mbox{\tiny GS}} \vert \Psi^{\mbox{\tiny DEX}}_{\mbox{\tiny DEX}} \rangle \\
  \langle \Psi^{\mbox{\tiny ENS}}_{\mbox{\tiny OSS}} \vert \Psi^{\mbox{\tiny GS}}_{\mbox{\tiny GS}} \rangle &
  \langle \Psi^{\mbox{\tiny ENS}}_{\mbox{\tiny OSS}} \vert \Psi^{\mbox{\tiny ENS}}_{\mbox{\tiny OSS}} \rangle &
  \langle \Psi^{\mbox{\tiny ENS}}_{\mbox{\tiny OSS}} \vert \Psi^{\mbox{\tiny DEX}}_{\mbox{\tiny DEX}} \rangle \\
  \langle \Psi^{\mbox{\tiny DEX}}_{\mbox{\tiny DEX}} \vert \Psi^{\mbox{\tiny GS}}_{\mbox{\tiny GS}} \rangle &
  \langle \Psi^{\mbox{\tiny DEX}}_{\mbox{\tiny DEX}} \vert \Psi^{\mbox{\tiny ENS}}_{\mbox{\tiny OSS}} \rangle &
  \langle \Psi^{\mbox{\tiny DEX}}_{\mbox{\tiny DEX}} \vert \Psi^{\mbox{\tiny DEX}}_{\mbox{\tiny DEX}} \rangle 
  \end{array}
  \right] \nonumber \\
  & = & \left[
  \begin{array}{ccc}
  1 & \sqrt{2} \langle i^{\mbox{\tiny GS}} \vert i^{\mbox{\tiny ENS}} \rangle 
  \langle i^{\mbox{\tiny GS}} \vert a^{\mbox{\tiny ENS}} \rangle &
  \langle i^{\mbox{\tiny GS}} \vert a^{\mbox{\tiny DEX}} \rangle^2 \\
  \sqrt{2} \langle i^{\mbox{\tiny ENS}} \vert i^{\mbox{\tiny GS}} \rangle \langle a^{\mbox{\tiny ENS}} \vert i^{\mbox{\tiny GS}} \rangle &
  1 &
  \sqrt{2} \langle i^{\mbox{\tiny ENS}} \vert a^{\mbox{\tiny DEX}} \rangle \langle a^{\mbox{\tiny ENS}} \vert a^{\mbox{\tiny DEX}} \rangle \\
  \langle a^{\mbox{\tiny DEX}} \vert i^{\mbox{\tiny GS}} \rangle^2 &
  \sqrt{2} \langle a^{\mbox{\tiny DEX}} \vert i^{\mbox{\tiny ENS}} \rangle 
  \langle a^{\mbox{\tiny DEX}} \vert a^{\mbox{\tiny ENS}} \rangle & 1  
  \end{array}
  \right] \, . \nonumber \\
  & & \label{eq:theory.23}
\end{eqnarray}
Those familiar with VB theory know that the hamiltonian matrix 
elements will also include orbital overlaps.  That is, we must
solve a new problem,
\begin{equation}
  {\bf H}' \vec{C}_I = E_I {\bf S} \vec{C}_I \, , 
  \label{eq:theory.12}
\end{equation}
where in WFT,
\begin{equation}
  {\bf H}' = \left[ \begin{array}{ccc} E^{\mbox{\tiny GS}}_0 & \sqrt{2}D & B \\
                            \sqrt{2}D & E_M^{\mbox{\tiny ENS}} + A & \sqrt{2} C \\
                            B & \sqrt{2} C & E^{\mbox{\tiny DEX}}_D \end{array} \right]
  \, ,
  \label{eq:theory.12.1}
\end{equation}
with
\begin{eqnarray}
  A & = & (i_{\mbox{\tiny ENS}}, a_{\mbox{\tiny ENS}} \vert f_H \vert
       a_{\mbox{\tiny ENS}}, i_{\mbox{\tiny ENS}} ) \nonumber \\
  B & = & 2 \langle i_{\mbox{\tiny GS}} \vert \hat{h} \vert a_{\mbox{\tiny DEX}} \rangle
  \langle i_{\mbox{\tiny GS}} \vert a_{\mbox{\tiny DEX}} \rangle
  + ( i_{\mbox{\tiny GS}} , a_{\mbox{\tiny DEX}} \vert f_H \vert i_{\mbox{\tiny GS}} , a_{\mbox{\tiny DEX}} ) \nonumber \\
   C & = & \langle a_{\mbox{\tiny DEX}} \vert {\hat h} 
           \vert i_{\mbox{\tiny ENS}} \rangle
           \langle a_{\mbox{\tiny DEX}} \vert a_{\mbox{\tiny ENS}} \rangle
        +   \langle a_{\mbox{\tiny DEX}} \vert {\hat h} 
            \vert a_{\mbox{\tiny ENS}} \rangle 
            \langle a_{\mbox{\tiny DEX}} \vert i_{\mbox{\tiny ENS}} \rangle
        + (a_{\mbox{\tiny DEX}} , i_{\mbox{\tiny ENS}} \vert f_H \vert
           a_{\mbox{\tiny DEX}} , a_{\mbox{\tiny ENS}})
   \nonumber \\
  D & = & \langle i_{\mbox{\tiny GS}} \vert {\hat h} \vert a_{\mbox{\tiny ENS}} \rangle
  \langle i_{\mbox{\tiny GS}} \vert i_{\mbox{\tiny ENS}} \rangle 
  + \langle i_{\mbox{\tiny GS}} \vert {\hat h} \vert i_{\mbox{\tiny ENS}} \rangle
  \langle i_{\mbox{\tiny GS}} \vert a_{\mbox{\tiny ENS}} \rangle 
  + ( i_{\mbox{\tiny GS}} , a_{\mbox{\tiny ENS}} \vert f_H \vert i_{\mbox{\tiny GS}} i_{\mbox{\tiny ENS}} ) \, .
  \label{eq:theory.12.2}
\end{eqnarray}
Notice that ${\bf H}' \rightarrow {\bf H}$ in the limit of a single set 
of orthonormal reference orbitals because
\begin{eqnarray}
   A & \rightarrow & (i a \vert f_H \vert a i) \nonumber \\
   B & \rightarrow & ( i a \vert f_H \vert i a ) = A \mbox{  (if real orbitals)} 
   \nonumber \\
   C & \rightarrow & \langle a \vert {\hat h} \vert i \rangle
        + ( a i \vert f_H \vert a a ) = f_{a,i}[\gamma_{\mbox{\tiny DEX}}]
   \nonumber \\
   D & \rightarrow & \langle i \vert {\hat h} \vert a \rangle
      + ( i a \vert f_H \vert i i )  = f_{i,a}[\gamma_{\mbox{\tiny GS}}] \, ,
  \label{eq:theory.12.3}
\end{eqnarray}
where $f_H = 1/r_{1,2}$ is the Hartree kernel.  

Also notice how Eq.~(\ref{eq:theory.12.2}) is much more complicated than 
Eq.~(\ref{eq:theory.12.3}).  In the first place, it involves treating one-
and two-electron terms differently.  That adds an additional level of 
complexity that we want to avoid unless it turns out to be absolutely
necessary (which, of course, could be the case, but does not seem to be
so for LiH).  But the second reason is far more important: This is that
diag MSM DFT requires us to replace some of the terms with xc-terms
(energies or potentials) which will have to be evaluated at a specific
density.  The above formulae would lead to cross terms which should depend
upon {\em two} reference densities or upon a transition density between
two references.  Some people do take this approach \cite{LG2025} and we
find it very interesting.  But our goal is to keep the present model
as simple as possible.
Since 
we want to avoid (for this article) the evaluation of 
overly complicated matrix-element expressions, we will make an
{\em additional} approximation---namely that $\Phi^{\mbox{\tiny GS}}_{\mbox{\tiny GS}}$
and $\Phi^{\mbox{\tiny DEX}}_{\mbox{\tiny DEX}}$ may be expanded in terms of
ENS CSFs as,
\begin{eqnarray}
  \vec{\Psi}' & = & \vec{\Psi} {\bf M} \nonumber \\
  \vec{\Psi} & = & \vec{\Psi}' \left( {\bf M}^\dagger {\bf M}
  \right)^{-1} {\bf M}^\dagger \, ,
  \label{eq:theory.13}
\end{eqnarray}
where
\begin{eqnarray}
  \vec{\Psi} & = & \left( \begin{array}{ccc}
   \vert \Psi^{\mbox{\tiny ENS}}_{\mbox{\tiny GS}} \rangle &
   \vert \Psi^{\mbox{\tiny ENS}}_{\mbox{\tiny OSS}} \rangle &
   \vert \Psi^{\mbox{\tiny ENS}}_{\mbox{\tiny DEX}} \rangle \end{array} \right)
  \nonumber \\
  \vec{\Psi}' & = & \left( \begin{array}{ccc}
   \vert \Psi^{\mbox{\tiny GS}}_{\mbox{\tiny GS}} \rangle &
   \vert \Psi^{\mbox{\tiny ENS}}_{\mbox{\tiny OSS}} \rangle &
   \vert \Psi^{\mbox{\tiny DEX}}_{\mbox{\tiny DEX}} \end{array} \right)
   \label{eq:theory.14}
\end{eqnarray}
and 
\begin{equation}
  {\bf M} = \vec{\Psi}^\dagger \vec{\Psi}^\prime \, .
  \label{eq:theory.14b}
\end{equation}
That is, we assume that the multireference linear combination of relaxed
wave functions may be expanded as a linear combination of single reference
unrelaxed wave functions.  This results in a sort of projection of ${\bf H}$
onto ${\bf H}'$,
\begin{eqnarray}
  {\bf H} & = & \vec{\Psi}^\dagger {\hat H} \vec{\Psi}
  \nonumber \\
  {\bf H}' & = & ({\vec \Psi}')^\dagger {\hat H} \vec{\Psi}'
  = {\bf M}^\dagger {\bf H} {\bf M} \nonumber \\
  {\bf S}' & = & (\vec{\Psi}')^\dagger \vec{\Psi}'
  = {\bf M}^\dagger {\bf M} \, .
   \label{eq:theory.15}
\end{eqnarray}
Coherent with this approximation is that we now have a new overlap matrix
as well --- namely,
\begin{equation}
  {\bf S}' = (\vec{\Psi}')^\dagger \vec{\Psi}' = {\bf M}^\dagger {\bf M} \, .
  \label{eq:theory.15b}
\end{equation}
The result,
\begin{equation}
  {\bf M}^\dagger {\bf H} {\bf M} \vec{C}_I = 
  E_I {\bf M}^\dagger {\bf M} \vec{C}_I \, ,
  \label{eq:theory.15c}
\end{equation}
is rather curious because ${\bf M}^\dagger$ is invertible in our calculations
which leads to 
\begin{equation}
  {\bf H} \left( {\bf M} \vec{C}_I \right) = 
  E_I \left( {\bf M} \vec{C}_I \right) \, .
  \label{eq:theory.15d}
\end{equation}
Since this is practically the same as our original single-reference CI
equation [Eq.~(\ref{eq:theory.11})],  we may continue
to solve Eq.~(\ref{eq:theory.11}) just as in Articles {\bf I} and {\bf II},
except that {\em the matrix elements of} ${\bf H}$ {\em are evaluated 
using different relaxed reference states}!  Our wave function interpretation
will be based upon
\begin{equation}
  \vec{C}_I^\prime = {\bf M} \vec{C}_I \, .
  \label{eq:theory.15e}
\end{equation}

\begin{figure}  
\begin{center}
\includegraphics[width=0.80\textwidth]{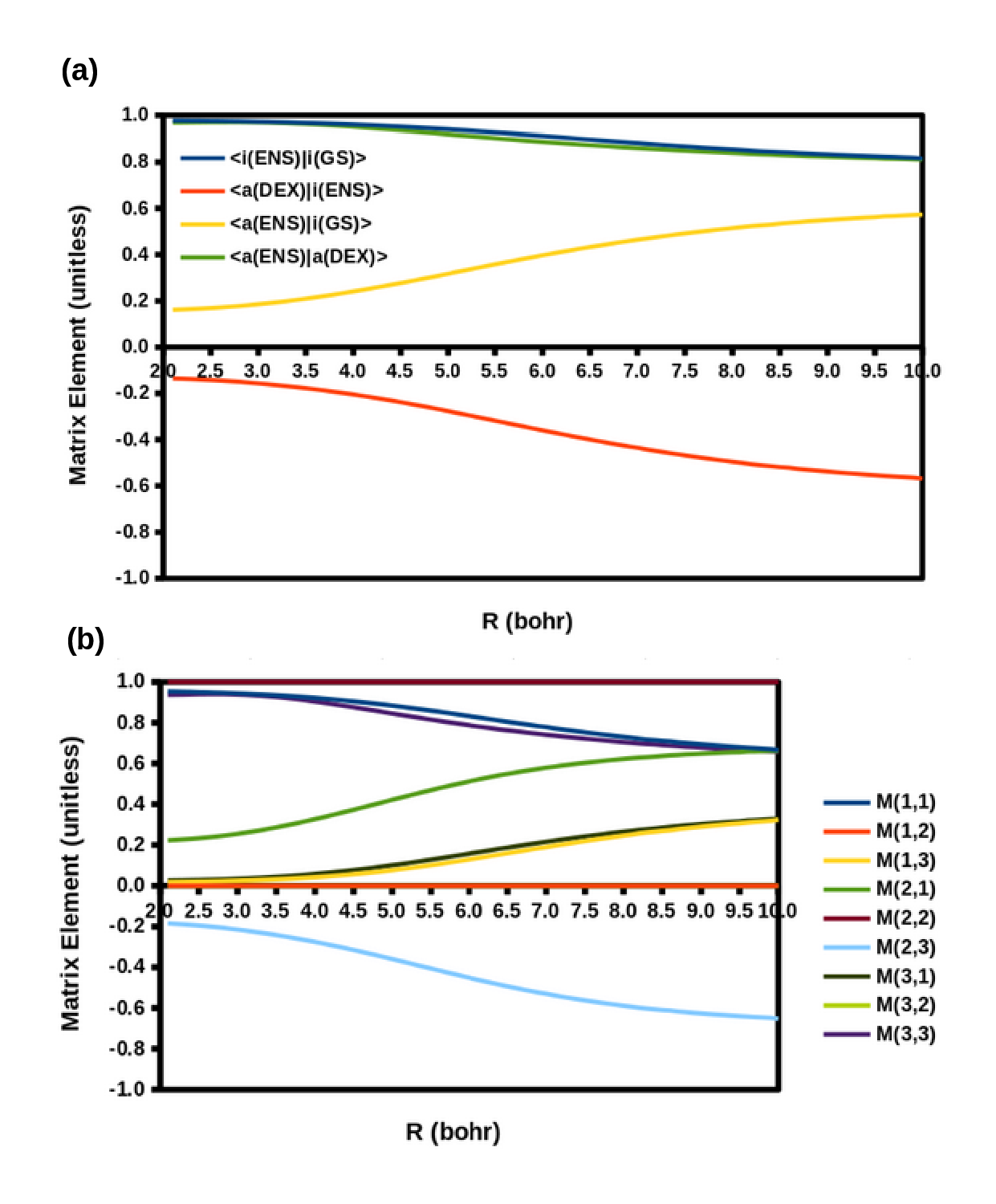}
\end{center}
\caption{
Matrix elements of the {\bf M} (b) and the overlap elements (a) used
to construct {\bf M}.  (See text.)
\label{fig:M}
}
\end{figure}
We have seen that ${\bf M}$ basically cancels out, but nevertheless enters
into the theory.  It can be calculated as 
\begin{eqnarray}
  {\bf M} & = & \left[
  \begin{array}{ccc}
  \langle \Psi^{\mbox{\tiny ENS}}_{\mbox{\tiny GS}} \vert \Psi^{\mbox{\tiny GS}}_{\mbox{\tiny GS}} \rangle &
  \langle \Psi^{\mbox{\tiny ENS}}_{\mbox{\tiny GS}} \vert \Psi^{\mbox{\tiny ENS}}_{\mbox{\tiny OSS}} \rangle &
  \langle \Psi^{\mbox{\tiny ENS}}_{\mbox{\tiny GS}} \vert \Psi^{\mbox{\tiny DEX}}_{\mbox{\tiny DEX}} \rangle \\
  \langle \Psi^{\mbox{\tiny ENS}}_{\mbox{\tiny OSS}} \vert \Psi^{\mbox{\tiny GS}}_{\mbox{\tiny GS}} \rangle &
  \langle \Psi^{\mbox{\tiny ENS}}_{\mbox{\tiny OSS}} \vert \Psi^{\mbox{\tiny ENS}}_{\mbox{\tiny OSS}} \rangle &
  \langle \Psi^{\mbox{\tiny ENS}}_{\mbox{\tiny OSS}} \vert \Psi^{\mbox{\tiny DEX}}_{\mbox{\tiny DEX}} \rangle \\
  \langle \Psi^{\mbox{\tiny ENS}}_{\mbox{\tiny DEX}} \vert \Psi^{\mbox{\tiny GS}}_{\mbox{\tiny GS}} \rangle &
  \langle \Psi^{\mbox{\tiny ENS}}_{\mbox{\tiny DEX}} \vert \Psi^{\mbox{\tiny ENS}}_{\mbox{\tiny OSS}} \rangle &
  \langle \Psi^{\mbox{\tiny ENS}}_{\mbox{\tiny DEX}} \vert \Psi^{\mbox{\tiny DEX}}_{\mbox{\tiny DEX}} \rangle 
  \end{array}
  \right] \nonumber \\
  & = & \left[ 
  \begin{array}{ccc}
  \langle i^{\mbox{\tiny GS}} \vert i^{\mbox{\tiny ENS}} \rangle^2 & 0 &
   \langle i^{\mbox{\tiny ENS}} \vert a^{\mbox{\tiny DEX}} \rangle^2 \\
  \sqrt{2} \langle i^{\mbox{\tiny ENS}} \vert i^{\mbox{\tiny GS}} \rangle 
  \langle a^{\mbox{\tiny ENS}} \vert i^{\mbox{\tiny GS}} \rangle
   & 1 &
   \sqrt{2} \langle i^{\mbox{\tiny ENS}} \vert a^{\mbox{\tiny DEX}} \rangle 
  \langle a^{\mbox{\tiny ENS}} \vert a^{\mbox{\tiny DEX}} \rangle
 \\
  \langle a^{\mbox{\tiny ENS}} \vert i^{\mbox{\tiny GS}} \rangle^2 & 0 &
   \langle a^{\mbox{\tiny ENS}} \vert a^{\mbox{\tiny DEX}} \rangle^2 
  \end{array}
  \right] \, .
  \label{eq:theory.20}
\end{eqnarray}
The matrix elements of ${\bf M}$ are shown in {\bf Fig.~\ref{fig:M}}.
Critically
$\langle a(\text{\tiny DEX}) \vert i(\text{\tiny ENS}) \rangle \approx 
-\langle a(\text{\tiny ENS}) \vert i(\text{\tiny GS}) \rangle$ which leads to
$M(2,3) \approx -M(2,1)$.  This will have consequences.

Of more interest is to calculate,
\begin{equation}
  {\bf S}' = \left[
  \begin{array}{ccc}
  \langle i^{\mbox{\tiny GS}} \vert {\hat P}^{\mbox{\tiny ENS}}_{\mbox{\tiny MO}} \vert i^{\mbox{\tiny GS}} \rangle^2 & 
  \sqrt{2} \langle i^{\mbox{\tiny GS}} \vert i^{\mbox{\tiny ENS}} \rangle \langle i^{\mbox{\tiny GS}} \vert a^{\mbox{\tiny ENS}} \rangle &
  \langle i^{\mbox{\tiny GS}} \vert {\hat P}^{\mbox{\tiny ENS}}_{\mbox{\tiny MO}} \vert a^{\mbox{\tiny DEX}} \rangle^2 \\
  \sqrt{2} \langle i^{\mbox{\tiny ENS}} \vert i^{\mbox{\tiny GS}} \rangle \langle a^{\mbox{\tiny ENS}} \vert i^{\mbox{\tiny GS}} \rangle &
  1 &
  \sqrt{2} \langle i^{\mbox{\tiny ENS}} \vert a^{\mbox{\tiny DEX}} \rangle \langle a^{\mbox{\tiny ENS}} \vert a^{\mbox{\tiny DEX}} \rangle \\
  \langle a^{\mbox{\tiny DEX}} \vert {\hat P}^{\mbox{\tiny ENS}}_{\mbox{\tiny MO}} \vert i^{\mbox{\tiny GS}} \rangle^2 &
  \sqrt{2} \langle a^{\mbox{\tiny DEX}} \vert i^{\mbox{\tiny ENS}} \rangle \langle a^{\mbox{\tiny DEX}} \vert a^{\mbox{\tiny ENS}} \rangle &
  \langle a^{\mbox{\tiny DEX}} \vert {\hat P}^{\mbox{\tiny ENS}}_{\mbox{\tiny MO}} \vert a^{\mbox{\tiny DEX}} \rangle^2 
  \end{array}
  \right] \, ,
  \label{eq:theory.21}
\end{equation}
whose relation to ${\bf S}$ is made clear by the use of 
\begin{equation}
  {\hat P}^{\mbox{\tiny ENS}}_{\mbox{\tiny MO}} = \vert i^{\mbox{\tiny ENS}} \rangle \langle i^{\mbox{\tiny ENS}} \vert
  + \vert a^{\mbox{\tiny ENS}} \rangle \langle a^{\mbox{\tiny ENS}} \vert \, ,
  \label{eq:theory.22}
\end{equation}
which is the projection onto the ENS MO space.  Replacing this projection 
operator with the identity operator gives the exact ${\bf S}$ 
[Eq.~(\ref{eq:theory.23})].

\begin{figure} 
\begin{center}
\includegraphics[width=\textwidth]{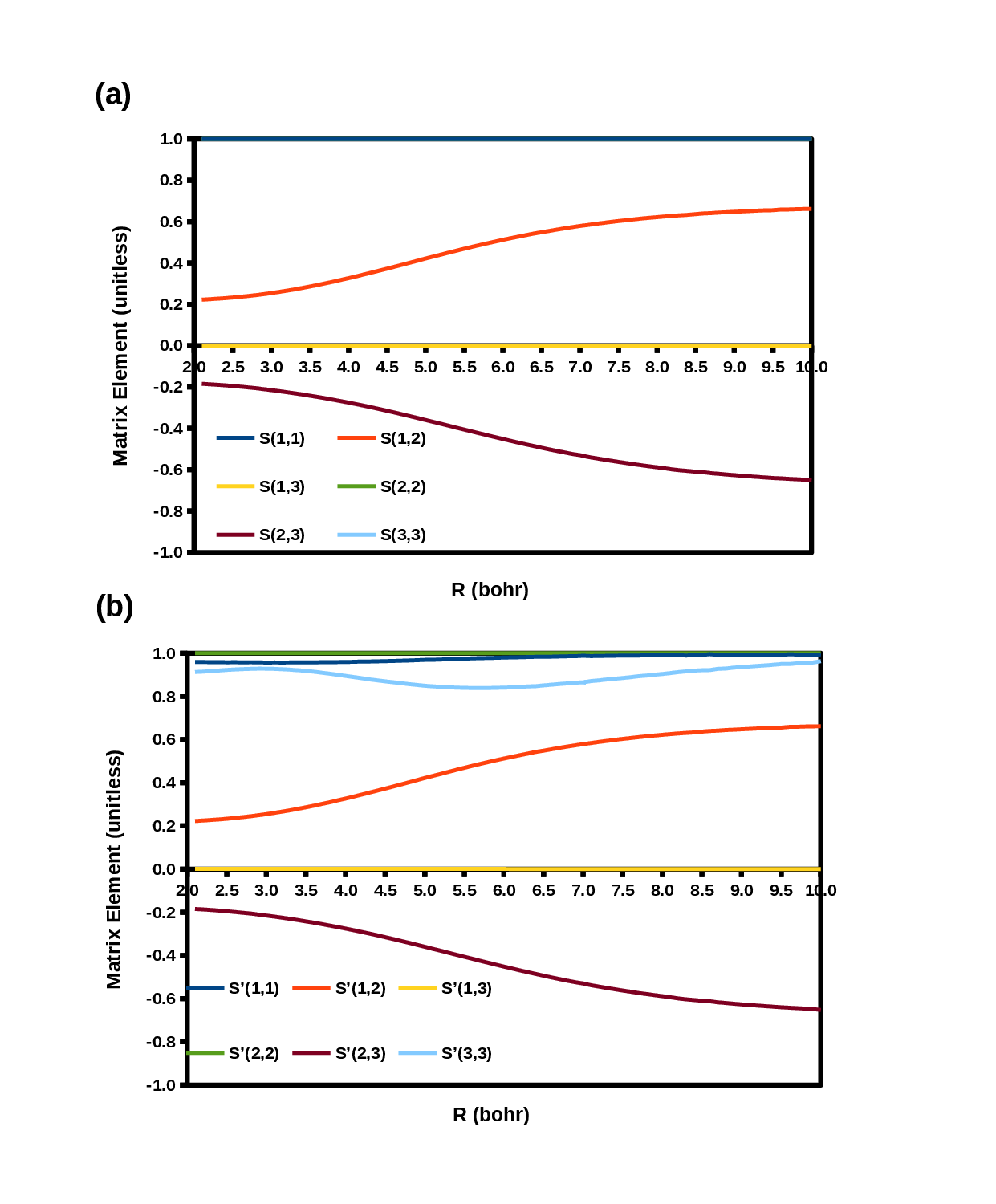} 
\end{center}
\caption{
Matrix elements of (a) {\bf S} and (b) {\bf S}' (see text).
\label{fig:S}
}
\end{figure}
{\bf Figure~\ref{fig:S}} provides a numerical comparison of the matrix elements
of ${\bf S}$ and ${\bf S}' = {\bf M}^\dagger {\bf M}$.  While there
are qualitative similarities, the most unexpected aspect is that
$S'(1,3) \approx 0$ which means that the GS and the DEX are effectively
decoupled in ${\bf S}'$ while they are coupled in ${\bf S}$.

\begin{figure} 
\begin{center}
\includegraphics[width=0.80\textwidth]{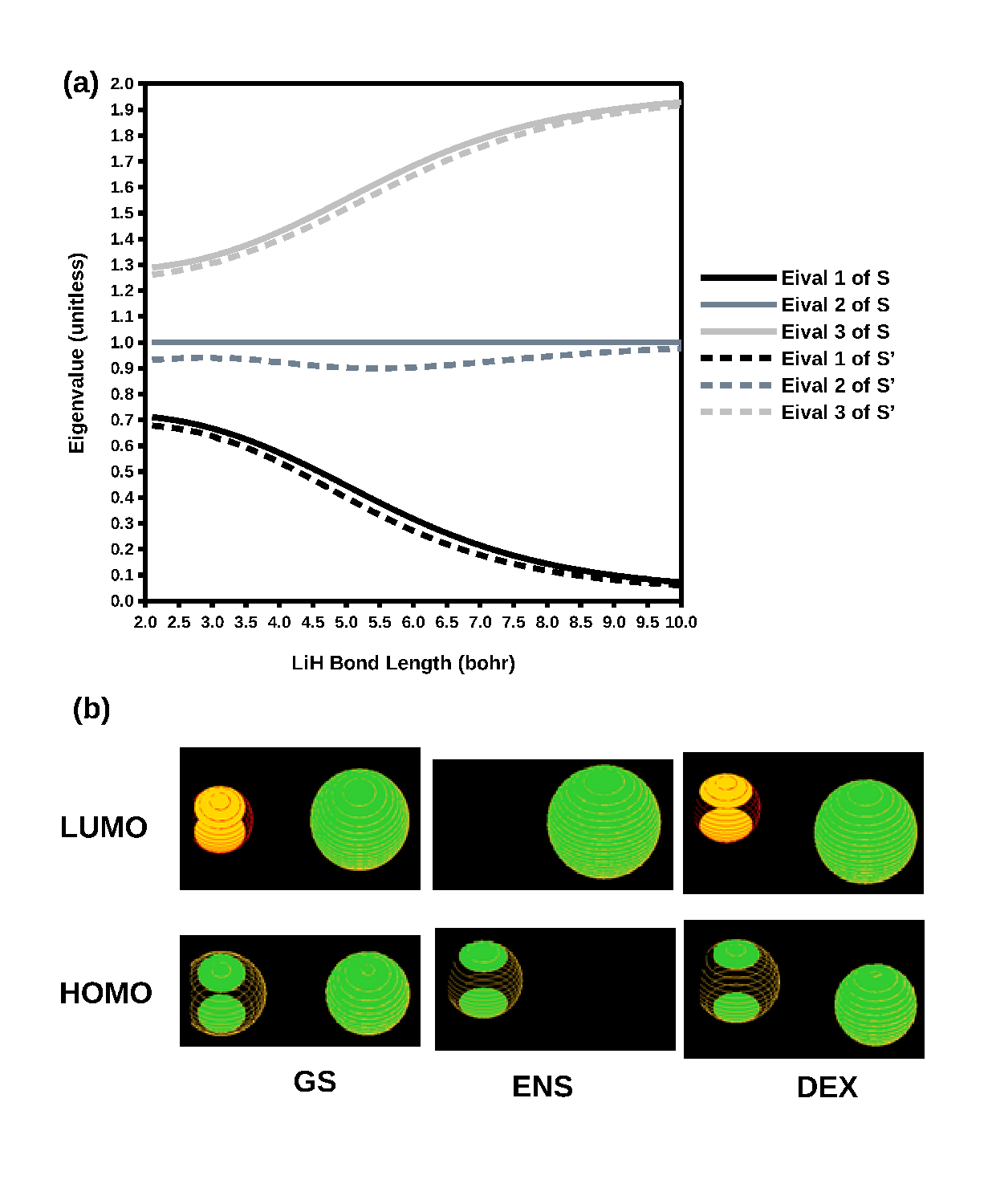} 
\end{center}
\caption{
(a) The eigenvalues of {\bf S} and {\bf S}' as a function of bond length.
(b) {\sc Molden} \cite{molden} images of the HOMO and LUMO calculated for 
the different references with a contour value of 0.03 a.u.  Note how the 
HOMO and LUMO of the GS and DEX SDETs are bonding and antibonding combinations
of similarly sized $s$-type orbitals while the HOMO of the ENS state is
an $s$-type orbital localized on H and the LUMO of the ENS state is an
$s$-type orbital localized on Li.
\label{fig:Seival}
}
\end{figure}
An even more important quantity is the lowest eigenvalue of the overlap
matrix.  The eigenvalues of ${\bf S}$ and of ${\bf S}'$ are shown in
{\bf Fig.~\ref{fig:Seival}(a)}.  The eigenvalues of ${\bf S}'$ are smaller
than, but in reasonable quantitative agreement with, those of ${\bf S}$ because
of the formulae for the matrix elements of ${\bf S}'$ contain a projection 
operator not found in the formulae for the matrix elements of ${\bf S}$.
The eigenvalues of an overlap matrix must always be positive, unless there 
are linear dependencies.  Such a linear dependency is developing at 10.0 bohr.
For ${\bf S}$, the MO coefficients read,
\begin{equation}
   0.504 \Psi_{\mbox{GS}} + -0.707 \Psi_{\mbox{OSS}} -0.496 \Psi_{\mbox{DEX}} \approx 0 \, ,
   \label{eq:depen.1}
\end{equation}
or
\begin{equation}
   \frac{1}{\sqrt{2}} \Psi_{\mbox{OSS}} \approx 
   \frac{1}{2} \left( \Psi_{\mbox{GS}} - \Psi_{\mbox{DEX}} \right) \, .
   \label{eq:depen.2}
\end{equation}
This apparently simple, and surprising, result is explained by the MOs
shown in Fig.~~\ref{fig:Seival}(b) which correspond roughly to,
\begin{eqnarray}
  \psi^{\mbox{GS}}_H & = & \frac{1}{\sqrt{2}} \left( \mbox{$1s$(H)} +
  \mbox{$2s$(Li)} \right) \nonumber \\
  \psi^{\mbox{OSS}}_H & = & \mbox{$1s$(H)} \nonumber \\
  \psi^{\mbox{OSS}}_L & = & \mbox{$2s$(Li)} \nonumber \\
  \psi^{\mbox{DEX}}_L & = & \frac{1}{\sqrt{2}} \left( \mbox{$1s$(H)} - \mbox{$2s$(Li)}
  \right) \, .
  \label{eq:depen.3}
\end{eqnarray}
From Table~\ref{tab:CSF}, Eq.~(\ref{eq:depen.2}) reads,
\begin{equation}
   \frac{1}{\sqrt{2}} \left( \vert H, \bar{L} \vert - \vert L, \bar{H} \vert 
   \right) \approx \frac{1}{2} \left( \vert H, \bar{H} \vert -
   \vert L , \bar{L} \vert \right) \, .
   \label{eq:depen.4}
\end{equation}
Inserting Eq.~(\ref{eq:depen.3}) and simplifying then gives,
\begin{equation}
  \frac{1}{\sqrt{2}} \left( \vert \mbox{$1s$(H)}, \overline{\mbox{$2s$(Li)}} \vert
  + \vert \mbox{$2s$(Li)} , \overline{\mbox{$1s$(H)}} \vert \right) \, ,
  \label{eq:depen.5}
\end{equation}
on both the left-hand and on the right-hand side of Eq.~(\ref{eq:depen.4}),
thus justifying the near linear dependence of the CSFs at 10.0 bohr.

A consequence of the semi-quantitative agreement between ${\bf S}$ and
${\bf S}'$ is ${\bf H} \vec{C}_I = E_I \vec{C}_I$, ${\bf H}' \vec{C}_I
= E_I {\bf S}' \vec{C}_I$, and ${\bf H}' \vec{C}_I = E_I {\bf S} \vec{C}_I$
should all produce very similar PECs.  This is true for 
${\bf H} \vec{C}_I = E_I \vec{C}_I$ and ${\bf H}' \vec{C}_I 
= E_I {\bf S}' \vec{C}_I$ because ${\bf H}' \vec{C}_I 
= E_I {\bf S}' \vec{C}_I$ is ${\bf M}^\dagger {\bf H} {\bf M} \vec{C}_I
= {\bf M}^\dagger {\bf M} \vec{C}_I$ and ${\bf M}^\dagger$ is invertible,
giving ${\bf H} \vec{C}^\prime_I = E_I \vec{C}^\prime_I$ with 
$\vec{C}^\prime_I = {\bf M} \vec{C}_I$.  It must also be approximately true for
${\bf H}' \vec{C}_I = E_I {\bf S} \vec{C}_I$ because ${\bf S} \approx {\bf S}'$.
Hence it suffices to diagonalize ${\bf H}^{\prime \prime} 
\vec{C}^{\prime \prime}_I = E_I \vec{C}^{\prime \prime}_I$ with 
${\bf H}^{\prime \prime} = {\bf S}^{-1/2} {\bf H}' {\bf S}^{-1/2}$ and
$\vec{C}^{\prime \prime} =  {\bf S}^{+1/2} \vec{C}_I$.
Our wavefunction analysis will be based directly upon $\vec{C}_I$, 
$\vec{C}^\prime_I$, or $\vec{C}^{\prime \prime}_I$, as appropriate, with
an additional Chirgwin-Coulson analysis (see Ref.~\cite{CC1950} and
the SI) for version {\bf v6}.

%
%

\section{Results and Analysis}
\label{sec:results}



\begin{table}
\begin{center}
\begin{tabular}{ccccccccc}
\hline \hline
Variation & \multicolumn{4}{c}{MOs} & \multicolumn{2}{c}{Coupling Matrices} & Hamiltonian & Overlap \\
{\bf v}         & GS & T & M & DEX & $C$ & $D$ &   \\
\hline
{\bf v0}  & ENS & ENS & ENS & ENS 
          & $f_{i,a}^{\mbox{\tiny ENS}}[\gamma_D^{\mbox{\tiny ENS}}]$ 
          & $f_{i,a}^{\mbox{\tiny ENS}}[\gamma_0^{\mbox{\tiny ENS}}]$ & ${\bf H}$ & {\bf 1} \\
{\bf v1}  & GS & GS & GS & GS 
          & $f_{i,a}^{\mbox{\tiny GS}}[\gamma_D^{\mbox{\tiny GS}}]$ 
          & 0 & ${\bf H}$ & {\bf 1} \\
({\bf v2})  & GS & GS & GS & GS 
          & $2 f_{i,a}^{\mbox{\tiny ENS}}[\gamma_D^{\mbox{\tiny ENS}}]$  
          & 0 & ${\bf H}$ & {\bf 1} \\
{\bf v3}  & GS & GS & GS & DEX
          & $2 f_{i,a}^{\mbox{\tiny ENS}}[\gamma_D^{\mbox{\tiny ENS}}]$ 
          & 0 & ${\bf H}$ & {\bf 1} \\
{\bf v4}  & GS & ENS & ENS & DEX
          & $f_{i,a}^{\mbox{\tiny ENS}}[\gamma_D^{\mbox{\tiny ENS}}]$
          & $f_{i,a}^{\mbox{\tiny ENS}}[\gamma_0^{\mbox{\tiny ENS}}] $ & ${\bf H}$ & {\bf 1} \\
{\bf v5}  & GS & ENS & ENS & DEX
          & $2 f_{i,a}^{\mbox{\tiny ENS}}[\gamma_D^{\mbox{\tiny ENS}}]$
          & 0 & ${\bf H}$ & {\bf 1} \\
{\bf v6}  & GS & ENS & ENS & DEX
          & $2 f_{i,a}^{\mbox{\tiny ENS}}[\gamma_D^{\mbox{\tiny ENS}}]$
          & 0 & ${\bf H}'$ & ${\bf S}$ \\
\hline \hline
\end{tabular}
\end{center}
\caption{\label{tab:variations} Variations tested: GS, groundstate; 
ENS, ensemble with half an electron of each spin in the HOMO and in the
LUMO; DEX, doubly excited.  Here ${\bf H}' = {\bf M}^\dagger {\bf H} {\bf M}$
and ${\bf S}' = {\bf M}^\dagger {\bf M}$.  The variation {\bf v2} is
in parentheses as this variation is only touched upon in the text. 
}
\end{table}


The aim of diag MSM DFT is to provide
a tool to help find missing matrix elements for a simple hybrid of DFT 
for dynamic correlation and CI for static and nondynamic correlation,  
but it is only a tool. 
The diagrams in diag MSM DFT provide a basis for
making educated guesses about how to calculate {\em appropriate} 
matrix elements in a small CI calculation.  
This is why the work reported here is exploratory in nature, just as it was in
Articles {\bf I} and {\bf II}. 
We are reminded of a popular quote often attributed to Thomas Edison that
reminds us that most scientific work necessarily involves a lot of trial
and error, notably in the early stages:
  ``I've tried everything. I have not failed. I have just found 10,000 
   ways that won't work!'' \cite{S83}
The present work is no exception.
Only a few of the variations tried (i.e., those shown in 
{\bf Table~\ref{tab:variations}}) 
will be discussed in order to show the systematic evolution of the
theory through progressive inclusion of relaxation effects in NOCI.

\begin{figure}  
\begin{center}
\includegraphics[width=\textwidth]{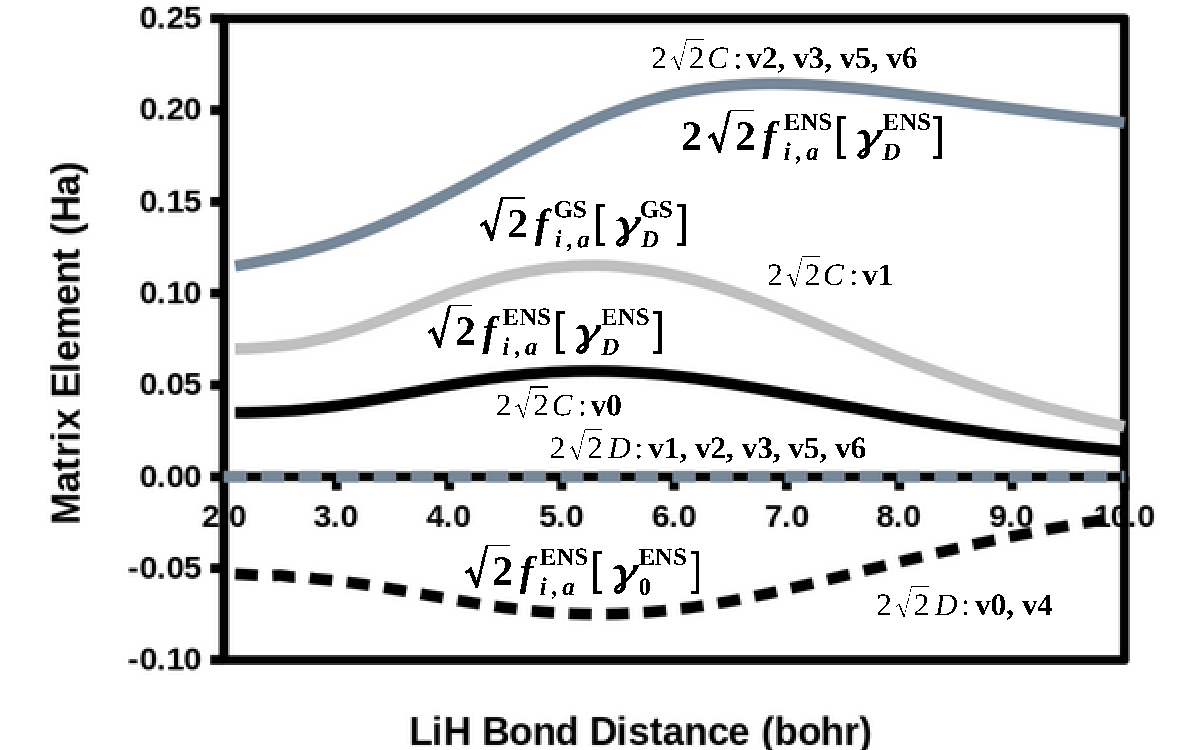} 
\end{center}
\caption{
The coupling matrix elements used in our study as a function of bond length.
(Only the variations actually discussed in the text have been indicated
in the graphic.)
\label{fig:LiH_LDA_Cv12}
}
\end{figure}
The coupling matrix elements in Table~\ref{tab:variations} are shown
in {\bf Fig.~\ref{fig:LiH_LDA_Cv12}}.  They are particularly important.
Matrix element $D$ is the direct coupling between the GS CSF and the OSS CSF
(GS $\leftrightarrow$ OSS).
For the variations discussed here, $D \neq 0$ only for {\bf v0} and 
{\bf v4}. 
All other variations use GS MOs to describe the GS, so imposing
``Brillouin's theorem'' (i.e., $D=0$) makes sense.  This means that there is  
no {\em direct} coupling between the GS SDET and the 
OSS SDET (GS $\leftarrow\negthickspace \negthickspace /\negthickspace \negthickspace \rightarrow$ OSS),
the coupling must pass through the DEX SDET
(GS $\longleftrightarrow$ DEX $\longleftrightarrow$ OSS).  
Therefore the magnitude of the $C$ matrix element is particularly important.
The variations discussed here with $D=0$ are {\bf v1}, {\bf v2}, {\bf v3},
{\bf v5}, and {\bf v6}.  
It is also a beautiful and simplifying feature of the current theory that,
except for {\bf v1}, all the off-diagonal NOCI matrix elements are calculated
with the ENS MOs as the ENS reference plays a central part in our theory.
This is also not quite true in {\bf v6} where matrix elements of ${\bf M}$
and ${\bf S}$ involve overlaps between GS, ENS, and DEX orbitals, but we will
show that the PECs of {\bf v6} and {\bf v5} are nearly indistinguishable. 

\begin{figure}  
\begin{center}
\includegraphics[width=0.60\textwidth]{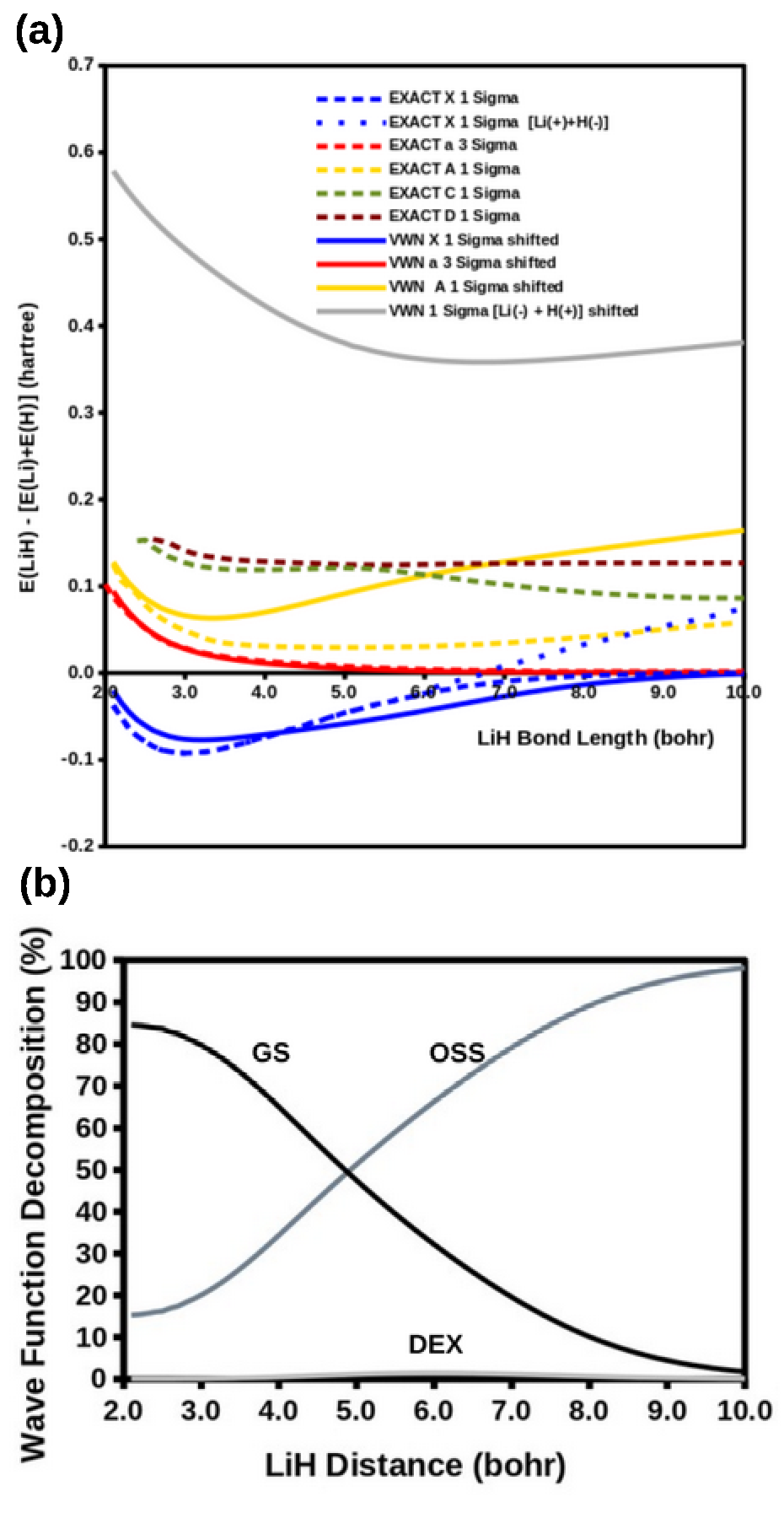} 
\end{center}
\caption{
(a) Comparison of EXACT and MSM-LDA ({\bf v0}) PECs with the energy shifted so 
that the $a \,^3\Sigma$ curve goes to zero at $R = \mbox{10.0 bohr}$.
(b) Wave-function decomposition of the {\bf v0} MSM-LDA ground state.
\label{fig:LiH_LDA_PECs}
}
\end{figure}
{\bf Figure~\ref{fig:LiH_LDA_PECs}} (a) shows how the excited-state 
MSM-LDA PECs of Article {\bf II} ({\bf v0}) compare with the EXACT PECs.  
Notice that the MSM-LDA triplet PEC is nearly superimposed with the 
$a \,^3\Sigma$ EXACT PEC.  The MSM-LDA and EXACT $A \,^1\Sigma$ PECs
are in good agreement at small bond distances but the MSM-LDA $A \,^1\Sigma$
PEC overestimates the EXACT $A \,^1\Sigma$ PEC at larger bond distances, 
consistent with the idea already mentioned in connection with the
GS curve that the coupling matrix element is too large at longer
bond distances.  There is no EXACT [Li:$^-$  H$^+$] PEC to compare with our 
MSM-LDA [Li:$^-$  H$^+$] PEC.
By construction, the MSM-LDA states are free of spin contamination and
dissociate correctly.  Nevertheless, the use of ENS MOs for describing
the ground state PEC results in underbinding.  Furthermore, the 
GS PEC has a peculiar shape which may be associated with an 
overestimation of the coupling matrix elements between the diabatic
states.
Figure~\ref{fig:LiH_LDA_PECs} (b) provides a wave-function decomposition
of the {\bf v0} $X\,^1\Sigma$ wave function.  As expected, the character
of the $X\,^1\Sigma$ changes from the ionic GS CSF to the biradical OSS CSF 
as the LiH bond is stretched.  As the non-zero $D$ matrix element provides
a direct coupling between the GS and OSS CSFs, configuration mixing with the
DEX configuration appears to be negligeable.

Of course, the GS MOs are variationally optimized for the lowest state,
meaning that the use of ENS MOs to construct the GS CSF is questionable.
If we follow the analogy of conventional CI, we would normally construct
all of the CSFs from the GS MOs.  This is {\bf v1}.
For the $C$ and $D$ matrix elements, we use 
\begin{eqnarray}
  C & = & f^{\mbox{\tiny GS}}_{i,a}[\gamma_D^{\mbox{\tiny GS}}] \nonumber \\
  D & = & f^{\mbox{\tiny GS}}_{i,a}[\gamma_0^{\mbox{\tiny GS}}] = 0 \, ,
  \label{eq:results.1}
\end{eqnarray}
where $f^{\mbox{GS}}_{i,a}$ is the off-diagonal element of the KS orbital
hamiltonian matrix calculated in the GS MO basis set.  
The expression for $D$ is the DFT analogue of Brillouin's theorem in 
wave-function theory.  

\begin{figure} 
\begin{center}
\includegraphics[width=0.60\textwidth]{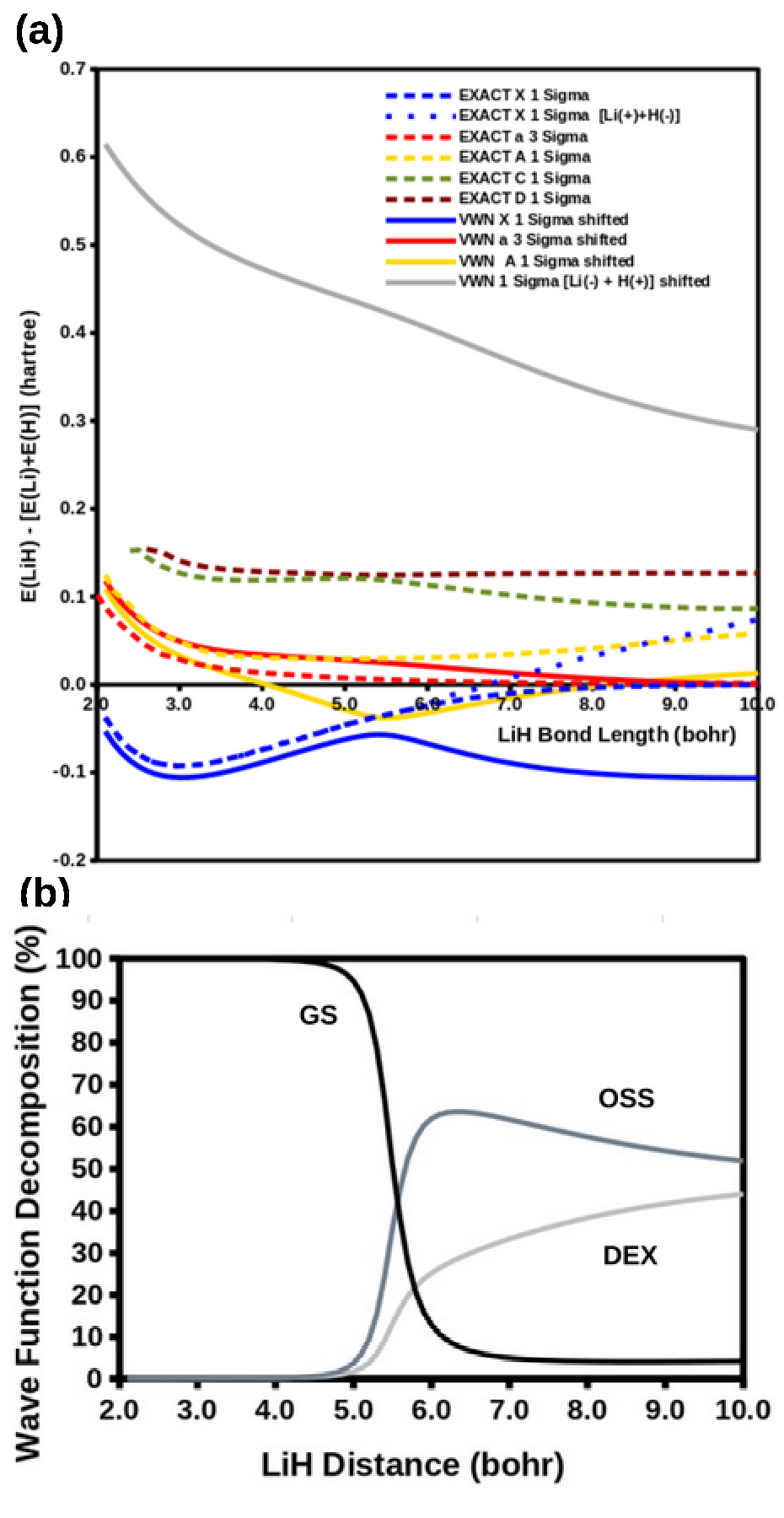} 
\end{center}
\caption{
(a) Comparison of EXACT and {\bf v1} MSM-LDA PECs with the energy shifted so 
that the $a \,^3\Sigma$ curve goes to zero at $R = \mbox{10.0 bohr}$.
(b) Wave-function decomposition of the {\bf v1} MSM-LDA ground state.
\label{fig:LiH_LDA_v1}
}
\end{figure}

{\bf Figure~\ref{fig:LiH_LDA_v1}} (a) shows a comparison of the {\bf v1} 
MSM-LDA PECs with the EXACT PECs. 
The avoidance of the diabatic curves is much too large
in Fig.~\ref{fig:LiH_LDA_v1} (a), indicating that either $C$ is too large or
that the diabatic PECs are too close in energy, or both.
The wave function decomposition of the {\bf v1} diag MSM-LDA ground state is
shown in Fig.~\ref{fig:LiH_LDA_v1}(b). 
We emphasize that there
is no {\em direct} coupling between the GS SDET and the open-shell singles
(OSS) SDET (GS $\leftarrow\negthickspace \negthickspace /\negthickspace \negthickspace \rightarrow$ OSS) because $D=0$, 
so the coupling must pass through the doubly excited (DEX) SDET 
(GS $\longleftrightarrow$ DEX $\longleftrightarrow$ OSS).  However 
Fig.~\ref{fig:LiH_LDA_v1}(b) shows the expected result that the GS CSF
is important at the equilibrium geometry.  However we also expect the
OSS CSF to dominate at large $R$ without significant contribution from
the DEX CSF.  Instead, while 
Fig.~\ref{fig:LiH_LDA_v1} (b) shows that 
the largest contribution at large $R$ is from the OSS CSF, 
the DEX CSF contribution is very significant at large $R$, which is 
counter to chemical intuition that we should be generating an OSS diradical.
\begin{figure}  
\begin{center}
\includegraphics[width=0.60\textwidth]{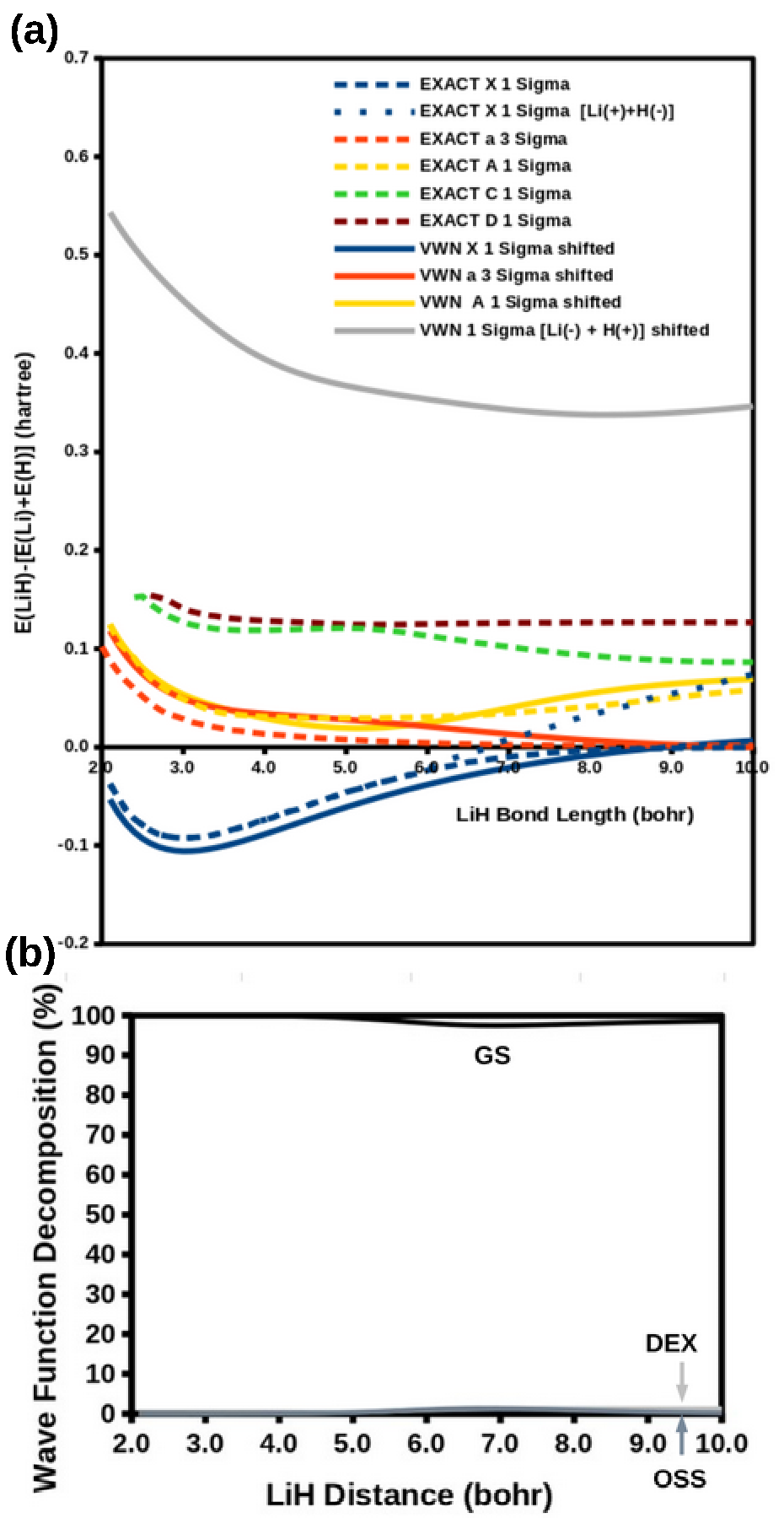}
\end{center}
\caption{
(a) Comparison of EXACT and {\bf v3} MSM-LDA PECs with the energy shifted so 
that the $a \,^3\Sigma$ curve goes to zero at $R = \mbox{10.0 bohr}$.
(b) Wave-function decomposition of the {\bf v3} MSM-LDA ground state.
\label{fig:LiH_LDA_v3}
}
\end{figure}

It was argued in Sec.~\ref{sec:theory} that the $C$ matrix element 
could also be approximated by using,
\begin{eqnarray}
  C & = & 2 f^{\mbox{\tiny ENS}}_{i,a}[\gamma_{\mbox{\tiny DEX}}^{\mbox{\tiny ENS}}] \nonumber \\
  D & = & 0 
  \label{eq:theory.9b}
\end{eqnarray}
[same as Eq.~(\ref{eq:theory.9})].  This gives us {\bf v2} whose 
PECs are not shown because it showed 
a collapse of the $^1\Sigma$ [Li$^-$ + H$^+$] PEC in that the energy 
is 0.14 Ha at $R = 10$ bohr, rather than the expected 0.35 Ha.  This caused
us to move onto {\bf v3} where the DEX determinant is calculated with DEX
MOs.  This fixed the problem because the [Li$^-$ + H$^+$] PEC now has the 
value of 0.35 Ha at $R = 10$ bohr, giving us the best overall results for 
the {\em singlet} curves as shown in {\bf Fig.~\ref{fig:LiH_LDA_v3}} (a).  However
the {\em triplet} PEC has collapsed a bit, consistent with the previous
observation that calculating the triplet PEC with MSM DFT and the ENS reference
seems to be optimal.
Nevertheless the wave-function decomposition of {\bf v3} shown in 
Fig.~\ref{fig:LiH_LDA_v3} (b) is completely nonintuitive!

\begin{figure} 
\begin{center}
\includegraphics[width=0.60\textwidth]{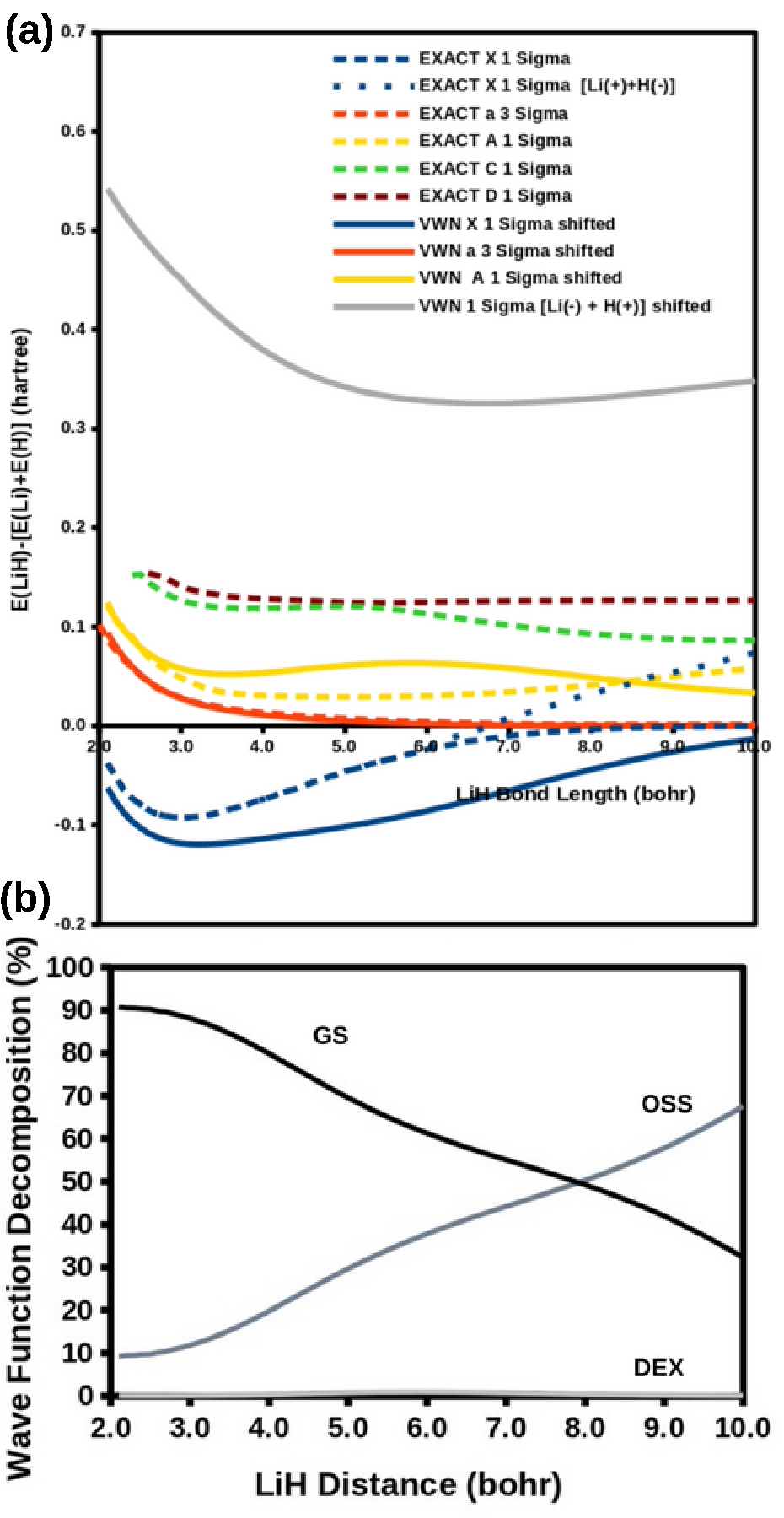} 
\end{center}
\caption{
(a) Comparison of EXACT and {\bf v4} MSM-LDA PECs with the energy shifted so 
that the $a \,^3\Sigma$ curve goes to zero at $R = \mbox{10.0 bohr}$.
(b) Wave-function decomposition of the {\bf v4} MSM-LDA ground state.
\label{fig:LiH_LDA_v4}
}
\end{figure}
Using the GS MOs to construct the GS CSF, the ENS MOs to construct
the triplet and OSS CSF, and the DEX MOs to construct the DEX CSF
gives {\bf v4}.  Comparing {\bf Fig.~\ref{fig:LiH_LDA_v4}} (a) with 
Fig.~\ref{fig:LiH_LDA_PECs} (a) , shows qualitative similarities though
{\bf v0} offers a better description of the $X\,^1\Sigma$ PEC
than does {\bf v4} and {\bf v4} offers a better description of the
$A\,^1\Sigma$ PEC than does {\bf v0}.  In fact, {\bf v4} offers a
much more convincing assignment of the calculated second excited
state with $A\,^1\Sigma$ than does {\bf v0}.  The wave-function
decomposition shown in 
Fig.~\ref{fig:LiH_LDA_v4} (b) is also
very similar to that of 
Fig.~\ref{fig:LiH_LDA_PECs} (b)

\begin{figure}  
\begin{center}
\includegraphics[width=0.60\textwidth]{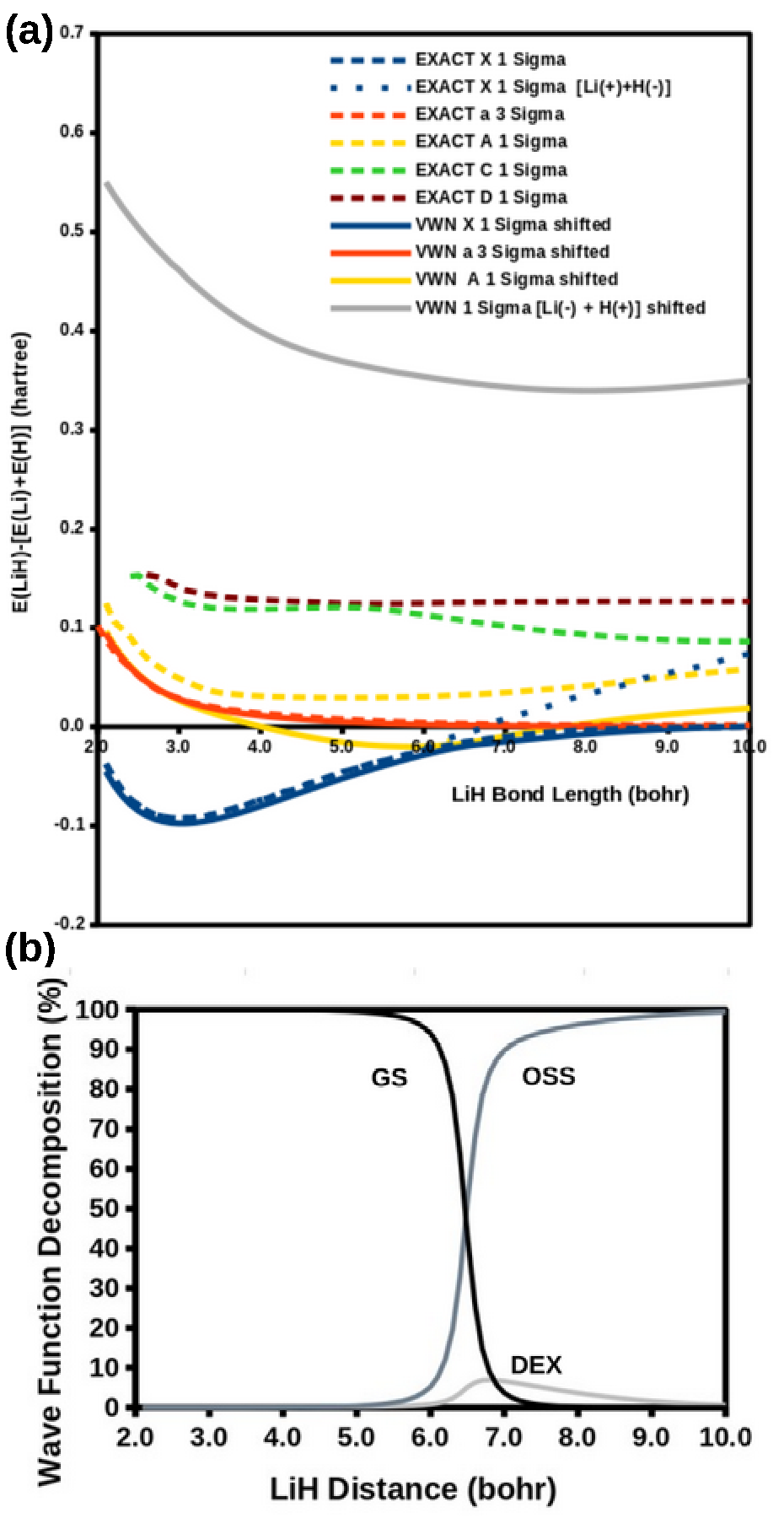}
\end{center}
\caption{
(a) Comparison of EXACT and {\bf v5} MSM-LDA PECs with the energy shifted so 
that the $a \,^3\Sigma$ curve goes to zero at $R = \mbox{10.0 bohr}$.
(b) Wave-function decomposition of the {\bf v5} MSM-LDA ground state.
\label{fig:LiH_LDA_v5}
}
\end{figure}
Of the other possibilities that we tried, {\bf v5} is the most
interesting and serendipitous variant.  It is identical to {\bf v4}
except that we used Eq.~(\ref{eq:theory.9b}) rather than
\begin{eqnarray}
  C & = &  f^{\mbox{\tiny ENS}}_{i,a}[\gamma_{\mbox{\tiny DEX}}^{\mbox{\tiny ENS}}] \nonumber \\
  D & = &  f^{\mbox{\tiny ENS}}_{i,a}[\gamma_{\mbox{\tiny GS}}^{\mbox{\tiny ENS}}]
  \label{eq:theory.8b}
\end{eqnarray}
[same as Eq.~(\ref{eq:theory.8})].  As shown in 
{\bf Fig.~\ref{fig:LiH_LDA_v5}} (a) , there is now a marked improvement in the
calculated $X\,^1\Sigma$ PEC.  Furthermore, the wave-function decomposition
of this state shown in  
Fig.~\ref{fig:LiH_LDA_v5} (b) is much more physical.
That is, it begins dominated by the GS CSF at the equilibrium geometry and
then switches to being dominated by the OSS CSF at large $R$.  Since coupling
between the GS CSF and OSS CSF is indirect and must pass through the DEX
CSF, we expect some contribution from the DEX CSF in the avoided crossing
region and this is exactly what we see.

\begin{figure}  
\begin{center}
\includegraphics[width=\textwidth]{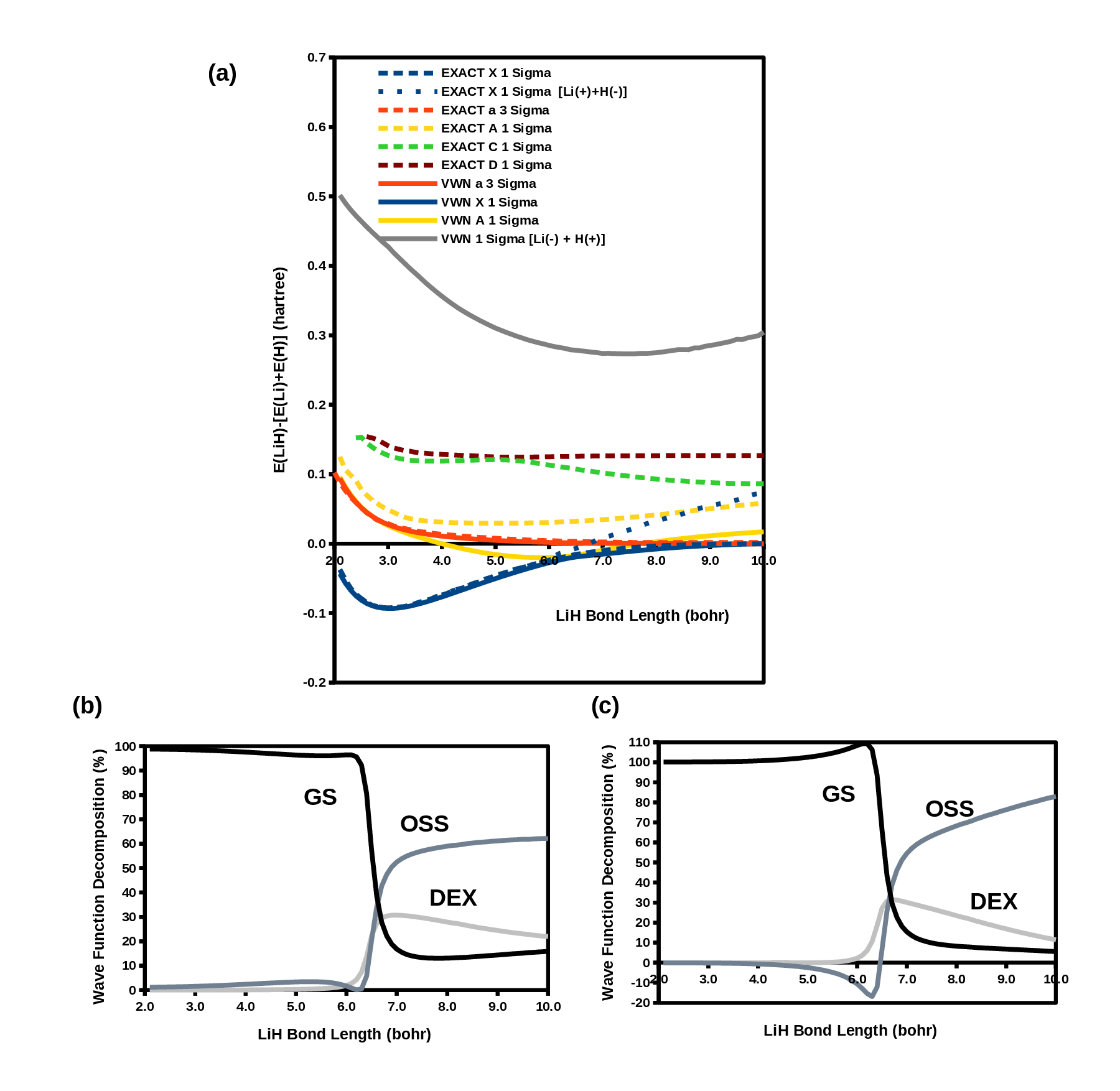}
\end{center}
\caption{
(a) Comparison of EXACT and {\bf v6} MSM-LDA PECs with the energy shifted so 
that the $a \,^3\Sigma$ curve goes to zero at $R = \mbox{10.0 bohr}$.
Wave-function decomposition of the {\bf v6} MSM-LDA ground state:
(b) $\vert C_{\mbox{GS}}^{\prime \prime}\vert^2 \times$ 100\%, 
(c) Chirgwin-Coulson analysis.
\label{fig:LiH_LDA_v6}
}
\end{figure}
Our initial intent was to do NOCI by solving 
${\bf H}' \vec{C}_I = E {\bf S}' \vec{C}_I$.  However we soon realized
that this gives us the same PECs as does solving 
${\bf H} \vec{C}_I^\prime = E \vec{C}_I^\prime$ because ${\bf M}^\dagger$
is invertible.  But we also showed that ${\bf S}' \approx {\bf S}$, so it
is interesting to check what happens if we replace ${\bf S}'$ with the
more exact ${\bf S}$ and solved ${\bf H}' \vec{C}_I = E {\bf S} \vec{C}_I$.
This is what we call {\bf v6}.  As comparison of 
{\bf Fig.~\ref{fig:LiH_LDA_v6}}(a) with Fig.~\ref{fig:LiH_LDA_v5}(a)
shows, there is no noticable difference in PECs between {\bf v5} and {\bf v6}.
Closer comparison of the two GS PECs with the EXACT GS PEC (not shown) reveals
{\bf v5} and {\bf v6} give nearly identical results, but
that {\bf v6} is ever so slightly worse than {\bf v5}, so we recommend {\bf v5} 
is preferred for accurate calculations and for the simplicity and 
internal consistency of the model.

In contrast to the computed PECs, the wave function decomposition shown
in Figs.~\ref{fig:LiH_LDA_v5}(b) and \ref{fig:LiH_LDA_v6}(b) and (c)
do show significant differences.  Of course, these differences are only
important for interpretation, but the interpretation of the curves is part
of what has been driving our development of diag MSM DFT.  In comparing
Fig.~\ref{fig:LiH_LDA_v5}(b) and Fig.~\ref{fig:LiH_LDA_v6}(b), we
are comparing $\vec{C}_{\text{GS}}^\prime = {\bf M} \vec{C}_{\text{GS}}$
with $\vec{C}_{text{GS}}^{\prime \prime} = {\bf S}^{+1/2} \vec{C}_{\text{GS}}$.
The replacement of ${\bf M}$ with ${\bf S}^{+1/2}$ is difficult to justify, but
there are qualitative similarities between the graphs.  A better comparison 
is between Fig.~\ref{fig:LiH_LDA_v5}(b) and the Chirgwin-Coulson analysis
shown in Fig.~\ref{fig:LiH_LDA_v6}(c).  Chirgwin-Coulson analysis
\cite{CC1950} was specifically developed for NOCI.  This type of analysis
is briefly reviewed in the SI where it may be seen to be very similar to
Mulliken population analysis.  While it has the advantage of ease of 
application, it shares the problem of Mulliken population analysis that
component populations can become negative while others exceed 100\% in 
order to maintain a total population of 100\%.  This is clearly seen in
Fig.~\ref{fig:LiH_LDA_v6}(c).  Nevertheless, Fig.~\ref{fig:LiH_LDA_v6}(c)
is in better qualitative agreement with Fig.~\ref{fig:LiH_LDA_v5}(b) than
is Fig.~\ref{fig:LiH_LDA_v6}(b).

\begin{figure} 
\begin{center}
\includegraphics[width=0.80\textwidth]{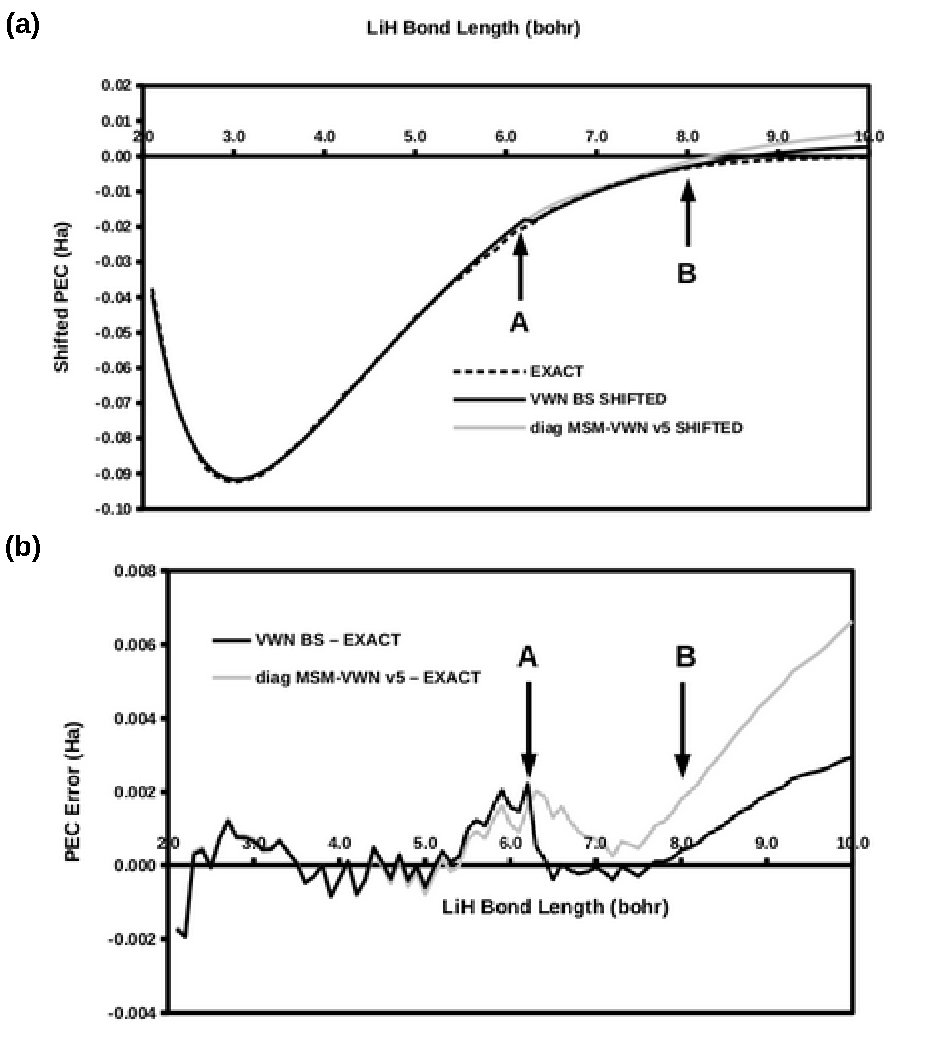} 
\end{center}
\caption{
Comparison of BS and {\bf v5} MSM-LDA GS shifted PECs with the EXACT GS PEC.
The BS and {\bf v5} MSM-LDA GS PECs have been shifted to have the same value
as the EXACT GS PEC at 3.8 bohr:
(a) EXACT, BS, and {\bf v5} MSM-LDA GS PECs and
(b) BS-EXACT and {\bf v5}-EXACT energy difference.
\label{fig:GS_PEC2}
}
\end{figure}

It is now useful to realize that the goal of NOCI is {\em not} typically
to obtain excited states, but rather to use excited states as a way to 
improve the ground-state PEC.  {\em To the extent that we accept this
point of view, we have succeeded in finding a relatively simple
SODS MDET DFT formalism which gives a ground-state PEC close to 
the usual broken-symmetry (BS) DODS DFT PEC without 
the spin-contamination and triplet instability associated with a 
broken-symmetry DODS calculation.}
This remarkable point is emphasized in {\bf Fig.~\ref{fig:GS_PEC2}}.  
Here the PECs have been shifted to agree with the EXACT GS PEC at 
3.8 bohr so that we may concentrate on the {\em shape} of the PECs.
Part (a) of the figure shows that the BS PEC is a little uneven at 
the Coulson-Fischer (symmetry-breaking) point located at distance {\bf A}. 
The calculations here are often a bit difficult to converge for technical
reasons.  In order to understand distance {\bf B}, the reader is referred back
to Fig.~\ref{fig:diabats}.  Beyond this distance it may be necessary
to include interactions with the Li($2p_z$)+H($1s$) diabatic curve in
order to get accurate results, but this is beyond the TOTEM that we
are using.  Par (b) of the figure is showing the difference between
the approximate and EXACT PECs.  Since the EXACT PECs were digitized,
the curves are a bit wiggly.  Neverthless it is clear that the two
approximate curves are nearly identical in shape up to the Coulson-Fischer
point {\bf A} and errors remain small (0.00200 Ha = 1.26 kcal/mol) until
point {\bf B} where errors increase for both approximate PECs.  After point 
{\bf B}, errors in {\bf v5} are larger than those in the BS calculation,
but they are still small and there is neither convergence difficulty nor
known slope discontinuities in the {\bf v5} approximation.  This is certainly
very encouraging since our long-range goal is to go on and do response
theory calculations of excited states that could be used in photochemical
dynamics.

We will not show results from the other variants that we tested.  They had
their place in building our understanding and in convincing ourselves that
we had at least tried the most obvious approaches.  Because of these other
calculations,  we have been able to identify and present only our most 
interesting and informative results.

%
%

\section{Concluding Discussion}
\label{sec:conclude}


In quantum chemistry, it is common to divide correlation into (``weak'') dynamic
correlation and ``strong'' correlation, where strong correlation is further
divided into static (``degenerate'') and nondynamic 
(``quasidegenerate'') correlation \cite{BS94}.  Most density-functional 
approximations (DFA) used in (approximate)
density-functional theory (DFT) include dynamic
correlation but not strong correlation. This means that practical DFT
calculations cannot describe spin- (and spatial-) multiplets nor can
they be expected to describe bond breaking and formation correctly as
these are processes where quasidegenerate single-determinant (SDET)
energies become quasidegenerate.  Hence there is a need for multideterminant
(MDET) DFT methods.  Especially valuable are MDET DFT methods which
require no reparameterization when using different DFAs and which avoid 
symmetry breaking (which can be problematic in response theory calculations).  
We have been slowly developing what we believe
to be a novel approach to MDET DFT which we call diagrammatic 
(diag) multiplet sum method (MSM) DFT.  As explained in
Article {\bf I}, diag MSM DFT is a hybrid DFT/WFT theory method in which a 
small configuration interaction (CI) matrix problem is constructed in which
off-diagonal CI matrix elements are obtained as educated guesses through 
comparison of the structure of CI matrix element (CIME) diagrams 
with DFT expansions.  A weakness of the method is that the educated guesses
must be tested numerically to see which are the most reliable.  However we
would like to believe that building up of this type of empirical experience
can lead to a reliable, parameter-free, same-orbital-for-different-spin 
(SODS) method for calculating the lowest energy potential energy
surface in chemical systems.  So far tests have been limited to the two-orbital
two-electron model (TOTEM) as applied to H$_2$, O$_2$, and LiH.
Article {\bf II} showed how charge transfer could be included in diag MSM
DFT in a simple fashion.  But the method failed to be quantitative for the
lowest energy state because it failed to include important molecular orbital
(MO) relaxation effects.  In this article, we have shown how 
relaxation effects may be incorporated into TOTEM calculations of the 
$X\,^1\Sigma$ potential energy curve (PEC) of LiH in such a way
that the resultant PEC is roughly as good as that obtained by using
broken symmetry (BS) with different-orbitals-for-different-spins
(DODS).  This is done using a small non-orthogonal CI 
(NOCI) calculation.  In this calculation, the ``excited states''
are fictitious in so far as they provide, at best, only approximate descriptions
of real excited states.  However they are key to improving the lowest energy
PEC.  We find these results highly promising and believe that our ideas can
help not only our efforts but also those of others working on developing
MDET DFT methods.  {\em We emphasize that, however well or badly this method
proves to work for other molecules, that it is already very interesting to
have found such a simple model that works so well for calculating an accurate
GS PEC for LiH.}

Of course, much more extensive testing on a variety of molecules will be
necessary before we can be truly confident that this is a robust method.  
So far, our work has been done almost entirely with a freely downloadable
version of the {\sc deMon2k} program on our laptop computers and a series
of simple home-made {\sc Python} codes.
(Only one small modification was made in the {\sc deMon2k} code in the
present work in order to guarantee SODS without any broken symmetry, although
this modification has only a tiny effect on the final results.)
The next step will be to automate the calculations within a private version
of the {\sc deMon2k} code so that we can test the method for more functionals,
basis sets, and other single $\sigma$ bonded diatomic molecules for which we 
expect the TOTEM to be valid.

%
%


\section*{Supplementary Information}
\label{sec:suppl}

The SI for this article contains:
\begin{enumerate}
\item Author Contributions
\item Computational Details
\item Origin of the Form of the $C$ and $D$ Matrix Elements
\item Chirgwin-Coulson analysis
\end{enumerate}

\section*{Acknowledgements}
\label{sec:thanks}


This is the continuation of a project initiated at the
African Schools on Electronic Structure Methods and Applications (ASESMA) 
that took place in Accra, Ghana, in 2025.  
We thank Lynn Scherwood Nkeme for helpful discussions about this project
at ASESMA 2025.  MEC thanks Kim C.\ Casida for helpful discussions
regarding the presentation of our results.



\end{document}